\documentclass[10pt]{article}
\usepackage{multicol}
\usepackage{graphicx}
\usepackage{amsmath}
\usepackage[a4paper]{geometry}
\usepackage{hyperref}

\begin{document}
\begin{center}
{\Large \bf Excitation function of initial temperature of heavy
flavor quarkonium emission source in high energy collisions}

\vskip.75cm

Qi Wang$^{1,2,}${\footnote{E-mail: 18303476022@163.com or
qiwang-sxu@qq.com}}, Fu-Hu Liu$^{1,2,}${\footnote{Correspondence
E-mail: fuhuliu@163.com or fuhuliu@sxu.edu.cn}}

\vskip.25cm

{\small\it $^1$Institute of Theoretical Physics \& State Key
Laboratory of Quantum Optics and Quantum Optics Devices,\\ Shanxi
University, Taiyuan, Shanxi 030006, People's Republic of China

$^2$Collaborative Innovation Center of Extreme Optics, Shanxi
University,\\ Taiyuan, Shanxi 030006, People's Republic of China}

\end{center}

\vskip.5cm

{\bf Abstract:} The transverse momentum spectra of $J/\psi$,
$\psi(2S)$, and $\Upsilon(nS, n=1,2,3)$ produced in proton-proton
($p$+$p$), proton-antiproton ($p$+$\overline{p}$), proton-lead
($p$+Pb), gold-gold (Au+Au), and lead-lead (Pb+Pb) collisions over
a wide energy range are analyzed by the (two-component) Erlang
distribution, the Hagedorn function (the inverse power-law), and
the Tsallis-Levy function. The initial temperature is obtained
from the color string percolation model from the fit by the
(two-component) Erlang distribution in the framework of
multisource thermal model. The excitation functions of several
parameters such as the mean transverse momentum and initial
temperature increase from 39 GeV to 13 TeV which is considered in
this work. The mean transverse momentum and initial temperature
decrease (increase slightly or do not change significantly) with
the increase of rapidity (centrality). Meanwhile, the mean
transverse momentum of $\Upsilon(nS, n=1,2,3)$ is larger than that
of $J/\psi$ and $\psi(2S)$, and the initial temperature for
$\Upsilon(nS, n=1,2,3)$ emission is higher than that for $J/\psi$
and $\psi(2S)$ emission, which shows a mass-dependent behavior.
\\

\\


\vskip1.0cm

\begin{multicols}{2}

{\section{Introduction}}

The excitation functions of some physical quantities are
significative to help us to understand the nuclear reaction
mechanism and the system evolution characteristic. For instance,
the higher the mean transverse momentum ($\langle p_T\rangle$) is,
the higher excitation state the emission source stays at.
Meanwhile, the higher the initial temperature
($T_i$)~\cite{1,1a,1aa,1aaa,1aaaa} is, the more violent the
collisions are. By the analysis of the excitation functions of
$\langle p_T\rangle$ and $T_i$, we can learn more about the
process in high energy collisions in which the excitation
functions of several parameters such as $\langle p_T\rangle$ and
$T_i$ can be obtained from the $p_T$ spectra of produced
particles.

In a data-driven reanalysis, to obtain $\langle p_T\rangle$ and
$T_i$, at the first place, we need the $p_T$ spectra of particles
in experiments. At the second place, we should choose appropriate
functions such as the Erlang distribution~\cite{1b,2,3}, the
Hagedorn function or the inverse power-law~\cite{4a,4b}, the
Tsallis-Levy function~\cite{5a,5b}, and others. At the last place,
we use the chosen functions to fit the experiential data on
particle spectra. By describing the $p_T$ spectra, the parameters
from the selected functions can be extracted. By comparing the
parameters obtained from the experiential data at different
energies, centralities, and rapidities, we can find out the
dependences of parameters on these quantities. These dependences
are related to excitation and expansion degrees of emission
source, which is beneficial for us to understand the mechanism and
characteristic of nuclear reactions and system evolution.

Besides the two derived parameters $\langle p_T\rangle$ and $T_i$,
we can obtain other related parameters by using the method which
is similar to extract $\langle p_T\rangle$ and $T_i$. For example,
using the Hagedorn function or the inverse power-law~\cite{4a,4b}
and the Tsallis-Levy function~\cite{5a,5b} to fit $p_T$ spectra,
some free parameters such as $p_0$, $n_0$, $T$, and $n$ in the
mentioned functions which will be discussed in section 2 can be
extracted. These free parameters are also useful to understand
particle productions and system evolution. Not only the excitation
functions of derived parameters $\langle p_T\rangle$ and $T_i$ but
also the trends of free parameters $p_0$, $n_0$, $T$, and $n$ can
be studied from the fit to $p_T$ spectra.

In this work, the (two-component) Erlang
distribution~\cite{1b,2,3}, Hagedorn function (the inverse
power-law)~\cite{4a,4b}, and Tsallis-Levy function~\cite{5a,5b}
are introduced firstly in section 2. Then, in section 3, the three
distributions or functions are used to preliminarily fit the $p_T$
spectra of heavy flavor quarkonia (charmonia and bottomonia)
produced in high energy collisions. The function results are
compared with the spectra of $J/\psi$, $\psi(2S)$, and
$\Upsilon(nS, n=1,2,3)$ measured by the STAR~\cite{27,28},
CDF~\cite{29a,29b,41}, ALICE~\cite{30},
LHCb~\cite{31,32,33,34,35,42,44,45,46}, ATLAS~\cite{36,38b,39},
and CMS Collaborations~\cite{37,38a,40,43} over a wide energy
range. Finally, in section 4, we give our summary and conclusions.
\\

{\section{Formalism and method}}

{\subsection{The (two-component) Erlang distribution}}

According to the multisource thermal model~\cite{1b,2,3}, a given
particle is produced in the collision process where a few partons
or quarks have taken part in. Each (the $i$-th) parton is assumed
to contribute to an exponential function [$f_i(p_t)$] of
transverse momentum ($p_t$) distribution. Let $\langle p_t
\rangle$ denotes the mean transverse momentum contributed by the
$i$-th parton, we have the probability density function of $p_t$
to be
\begin{align}
f_i(p_t)=\frac{1}{\langle p_t \rangle}
\exp\bigg(-\frac{p_t}{\langle p_t \rangle}\bigg)
\end{align}
which is normalized to 1. The probability density function of
$p_T$ contributed by all $N$ partons which have taken part in the
collision process is the convolution of $N$ exponential
functions~\cite{1b,2,3}. We have the $p_T$ distribution $f_E(p_T)$
(the probability density function of $p_T$) of final state
particles to be the Erlang distribution
\begin{align}
f_E(p_T)=\frac{p_T^{N-1}}{(N-1)!\langle p_t\rangle^{N}}
\exp\bigg(-\frac{p_T}{{\langle p_t\rangle}}\bigg)
\end{align}
which is naturally normalized to 1. The mean $p_T$ is $\langle
p_T\rangle=N \langle p_t\rangle$.

In the two-component Erlang distribution, we have
\begin{align}
f_1(p_T) &= \frac{k_Ep_T^{N_1-1}}{(N_1-1)!\langle
p_t\rangle_1^{N_1}}
\exp\bigg(-\frac{p_T}{{\langle p_t\rangle_1}}\bigg) \nonumber\\
&+ \frac{(1-k_E)p_T^{N_2-1}}{(N_2-1)!\langle p_t\rangle_2^{N_2}}
\exp\bigg(-\frac{p_T}{{\langle p_t\rangle_2}}\bigg),
\end{align}
where $k_E$ denotes the contribution fraction of the first
component, $N_1$ ($N_2$) denotes the number of partons in the
first (second) component, and $\langle p_t\rangle_1$ ($\langle
p_t\rangle_2$) denotes the mean transverse momentum contributed by
each parton in the first (second) component. The mean $p_T$ is
$\langle p_T\rangle=k_E N_1\langle p_t\rangle_1 +(1-k_E)N_2
\langle p_t\rangle_2$, where $N_1=1$--3 in this work and $N_2=2$
if $1-k_E\neq0$.
\\

{\subsection{The (two-component) Hagedorn function}}

The Hagedorn function is an inverse power-law which is suitable to
describe wide $p_T$ spectra of particles produced in hard
scattering process. In refs.~\cite{4a,4b}, the Hagedorn function
or the inverse power-law shows the probability density function of
$p_T$ to be
\begin{align}
f_H(p_T)=Ap_T\bigg(1+\frac{p_T}{p_0} \bigg)^{-n_0},
\end{align}
where $p_0$ and $n_0$ are the free parameters and $A$ is the
normalization constant which is related to $p_0$ and $n_0$ and
results in $\int_0^{\infty} f_H(p_T)dp_T=1$. Eq. (6) is an
empirical formula inspired by quantum chromodynamics (QCD). We
call Eq. (4) the Hagedorn function or the inverse
power-law~\cite{4a,4b}.

In the case of using two-component Hagedorn function, we have
\begin{align}
f_2(p_T)&= k_HA_1p_T\bigg(1+\frac{p_T}{p_{01}} \bigg)^{-n_{01}} \nonumber\\
&+ (1-k_H)A_2p_T\bigg(1+\frac{p_T}{p_{02}} \bigg)^{-n_{02}},
\end{align}
where $k_H$ denotes the contribution fraction of the first
component, $A_1$ ($A_2$) is the normalization constant which
results in the first (second) component to be normalized to 1, and
$p_{01}$ ($p_{02}$) and $n_{01}$ ($n_{02}$) are free parameters
related to the first (second) component. To combine the free
parameters of the two components, we have
$p_0=k_Hp_{01}+(1-k_H)p_{02}$ and $n_0=k_Hn_{01}+(1-k_H)n_{02}$.

Generally, Eq. (4) is possible to describe the spectra in both the
low- and high-$p_T$ regions. In fact, the spectra in the low- and
high-$p_T$ regions represent similar trend in some cases. This is
caused due to the similarity~\cite{6,7,8,9,10,11,12,13,14,15,16}
which is widely existent in high energy collisions, where the
similarity means the common or universality laws existed in
different processes or collisions. In addition, one can revise Eq.
(4) if needed in different ways~\cite{17,18,19,20,21,22,23} which
suppress in the spectrum itself in low- or high-$p_T$ region
according to the experimental spectra. To discuss various
revisions of the Hagedorn function or the inverse
power-law~\cite{4a,4b} is beyond the focus of this paper. We shall
not discuss anymore on this issue. For a very wide $p_T$ spectrum,
Eq. (5) is possibly needed.
\\

{\subsection{The (two-component) Tsallis-Levy function}}

The Tsallis statistics~\cite{5a} has wide applications in high
energy collisions. There are various forms of the Tsallis
distribution or function. In this work, we use the Tsallis-Levy
function~\cite{5b}
\begin{align}
f_L(p_T)=C p_T
\bigg(1+\frac{\sqrt{p_T^2+m_0^2}-m_0}{nT}\bigg)^{-n},
\end{align}
where $T$ and $n$ are free parameters, $\sqrt{p_T^2 + m_0^2}\equiv
m_T$ is the transverse mass, $m_0$ is the rest mass of the
considered particle, and $C$ is the normalized constant which is
related to $T$, $n$, and $m_0$ and results in $\int_0^{\infty}
f_L(p_T)dp_T=1$.

We notice that $f_L(p_T)$ is related to particle mass $m_0$, which
is not the case of $f_E(p_T)$ and $f_H(p_T)$ presented in Eqs. (2)
and (4) respectively. Although $f_L(p_T)$ is related to $m_0$,
this relation is not strong due to $m_0$ appearing only in
$\sqrt{p_T^2+m_0^2}-m_0$. The fact that the Tsallis distribution
depends on $m_0$ shows that this takes simple kinematics into
account, as it is well known that $m_T$ or $m_T-m_0$ (something
like transverse kinetic energy) is a better ``scaling variable"
for the spectra than $p_T$.

In the case of using two-component Tsallis-Levy function, we have
\begin{align}
f_3(p_T)&= k_LC_1 p_T
\bigg(1+\frac{\sqrt{p_T^2+m_0^2}-m_0}{n_1T_1}\bigg)^{-n_1} \nonumber\\
&+ (1-k_L)C_2 p_T \bigg(1+\frac{\sqrt{p_T^2+m_0^2}-m_0}
{n_2T_2}\bigg)^{-n_2},
\end{align}
where $k_L$ denotes the contribution fraction of the first
component, $C_1$ ($C_2$) is the normalization constant which
results in the first (second) component to be normalized to 1, and
$T_1$ ($T_2$) and $n_1$ ($n_2$) are free parameters. To combine
the free parameters of the two components, we have
$T=k_LT_1+(1-k_L)T_2$ and $n=k_Ln_1+(1-k_L)n_2$.

The temperature parameter in the Tsallis-Levy function is an
effective temperature at the final state (the stage of kinetic
freeze-out). This effective temperature is not a ``real"
temperature because it includes not only the contribution of
random thermal motion but also the contribution of flow effect. In
the case of the first (second) component having $T_1$ ($T_2$) with
the fraction of $k_L$ ($1-k_L$), the common effective temperature
$T$ of the two components is extracted from the assumed common
equilibrium state of the two components. That is
$T=k_LT_1+(1-k_L)T_2$ which has the same form as the parameter
$n$.
\\

{\subsection{The initial temperature}}

According to the color string percolation model~\cite{24,25,26},
the initial temperature of the emission source is determined by
\begin{align}
T_i \equiv\sqrt{\frac{\langle p_T^2 \rangle}{2}}
\end{align}
where
\begin{align}
\langle p_T^2\rangle=\int_0^{\infty} p_T^2 f_{1,2,3}(p_T)dp_T
\end{align}
is the square of the root-mean-square of $p_T$ due to
$\int_0^{\infty} f_{1,2,3}(p_T)dp_T=1$. If the $x$-component
($p_x$) and $y$-component ($p_y$) of the transverse momentum $p_T$
are considered, we have
\begin{align}
T_i =\sqrt{\langle p_x^2 \rangle}=\sqrt{\langle p_y^2 \rangle}.
\end{align}
In the source rest-frame and under the assumption of isotropic
emission, if the $z$-component of momentum is $p_z'$, we also have
\begin{align}
T_i =\sqrt{\langle p_z'^2 \rangle}.
\end{align}
Although the source rest-frame is the lab-frame for symmetric
collisions, we have mentioned the source rest-frame because
asymmetric proton-lead ($p$+Pb) collisions are also considered in
this work.

It should be noted that we have used a single string in the
cluster for a given particle production because only a projectile
participant quark and a target participant quark are mainly
considered in our treatment. The assumption of the single string
results in the color suppression factor $F(\xi)$ to be 1 in the
color string percolation model~\cite{25}. If we consider more than
one strings taking part in the given particle production, the
minimum $F(\xi)$ will be nearly 0.6~\cite{25}. Thus, we shall
obtain a higher $T_i$ by multiplying a revised factor
$1/\sqrt{F(\xi)}$ in Eqs. (8), (10), and (11). In our opinion,
although more than one strings have influences on the given
particle production, the main role is the single string.
\\

{\subsection{Discussion on the functions}}

We would like to point out that the three types of functions are
mainly just used here as parametrizations to achieve a good fit to
the data, to be able to extract $\langle p_T\rangle$ and $T_i$,
though the Hagedorn and Tsallis-Levy functions are physically
relevant. In fact, in the two functions, if we let $m_0=0$,
$p_0=nT$, $n_0=n=1/(q-1)$, the two functions are the same. Here
$q$ is an entropy index that characterizes the excitation degree
of the collision system~\cite{5a,5b}. Generally, $n_0$ or $n$ is a
sizeable quantity, which results in $q$ to close to 1 and the
collision system to close to an equilibrium state.

We have used the two-component functions in some cases. The reason
for using two-component source, i.e. basically two temperatures is
not just used to achieve a better fit to the data. Physically, the
first component corresponds to the non-head-on collisions between
projectile and target participant quarks. The second component
corresponds to the head-on collisions between the two quarks.
Generally, the first component has a large fraction and low
$\langle p_T\rangle$ and $T_i$. The second component has a less
fraction and high $\langle p_T\rangle$ and $T_i$. Because the
head-on collisions between the two quarks are infrequent, single
component function is usually applicable.

In principal, no matter what functions are used to fit the
experimental data, $\langle p_T\rangle$ (or $T_i$) obtained from
different fits are approximately the same within a small
systematic uncertainty, if different functions fit the data good
enough in the $p_T$ region of data available. For example, if
simple Maxwell-Boltzmann or Bose-Einstein statistics can fit the
data, we may obtain similar $\langle p_T\rangle$ (or $T_i$) with
other functions. In the case of multi-component Maxwell-Boltzmann
or Bose-Einstein statistics being needed, we may also obtain
similar $\langle p_T\rangle$ (or $T_i$).

Indeed, the data itself decides $\langle p_T\rangle$ (or $T_i$),
and $\langle p_T\rangle$ (or $T_i$) can be directly obtained from
the data itself. The reason why we use functions is to see the
tendency where the data is not available. However, the
extrapolation on the tendency should be careful because it is not
fully true, as it to the low- and high $p_T$ regions (where there
is no data) could in principle have a major effect on the tendency
(for example in case of very step exponentials near $p_T=0$, or
power-law tails at large $p_T$). To reduce the effect, the data
should be measured in a sufficiently large $p_T$ interval so that
the extrapolation does not spoil as far as possible.

For different components and functions, we do not need to consider
the values of mid-rapidity (mid-$y$) or mid-pseudorapidity
(mid-$\eta$), or the values of mid-$y$ or mid-$\eta$ can be
regarded as 0 directly. In fact, for the experimental data with
non-zero mid-$y$ or mid-$\eta$, we may directly regard them as
those with mid-$y=0$ or mid-$\eta=0$. This treatment is performed
to subtract the contribution of kinetic energy of directed motion
to the temperature.
\\

{\section{Results and discussion}}

Ordered by center-of-mass energy per nucleon pair ($\sqrt{s_{NN}}$
or $\sqrt{s}$ if only one pair) for different panels, Fig. 1 shows
the $p_T$ spectra, (a)--(c) $(1/2\pi p_T)d^2N/dp_Tdy$, (d)(e)
$(1/2\pi p_T)d^2\sigma/dp_Tdy$, (f) $d^2\sigma/dp_Tdy$, and (g)
$d^2Y/dp_Tdy$, of (a)--(d) $J/\psi \rightarrow e^+e^-$, (e)
$J/\psi \rightarrow \mu^+\mu^-$, (f) prompt $J/\psi$, and (g)
inclusive $J/\psi$ produced in (a)--(c) gold-gold (Au+Au), (d)(e)
proton-proton ($p$+$p$), (f) proton-antiproton
($p$+$\overline{p}$), and (g) lead-lead (Pb+Pb) collisions at
mid-rapidity (a)--(d) $|y|<1$, (e) $|y|<0.4$, (f) $|y|<0.6$, and
forward rapidity (g) $2.5<y<4$ at $\sqrt{s_{NN}}$ or $\sqrt{s}=$
(a) 39, (b) 62.4, (c) 200, (d) 500, and (e) 510 GeV, as well as
(f) 1.8 and 1.96, and (g) 2.76 TeV, where $N$ denotes the number
of particles, $\sigma$ denotes the cross section, and $Y$ denotes
the yield. The symbols represent the experimental
data~\cite{27,28,29a,29b,30} and the curve are our fitted results.
In the calculations, the method of least square is used to obtain
the best free parameters. The values of free parameters $\langle
p_t\rangle_1$, $\langle p_t\rangle_2$, $N_1$, and $k_E$ are listed
in Table A1 with $\chi^2$ and number of degree of freedom (ndof)
in the appendix 1. The values of free parameters $p_0$, $n_0$,
$T$, and $n$ are listed in Table A2 with $\chi^2$ and ndof in the
appendix. One can see that the (two-component) Erlang
distribution, the Hagedorn function, and the Tsallis-Levy function
fit approximately the experimental $p_T$ spectra of $J/\psi$ via
different decay or production modes in high energy $p$+$p$,
$p$+$\overline{p}$, Au+Au, and Pb+Pb collisions.

\begin{figure*}[htbp]
\begin{center}
\includegraphics[width=9.5cm]{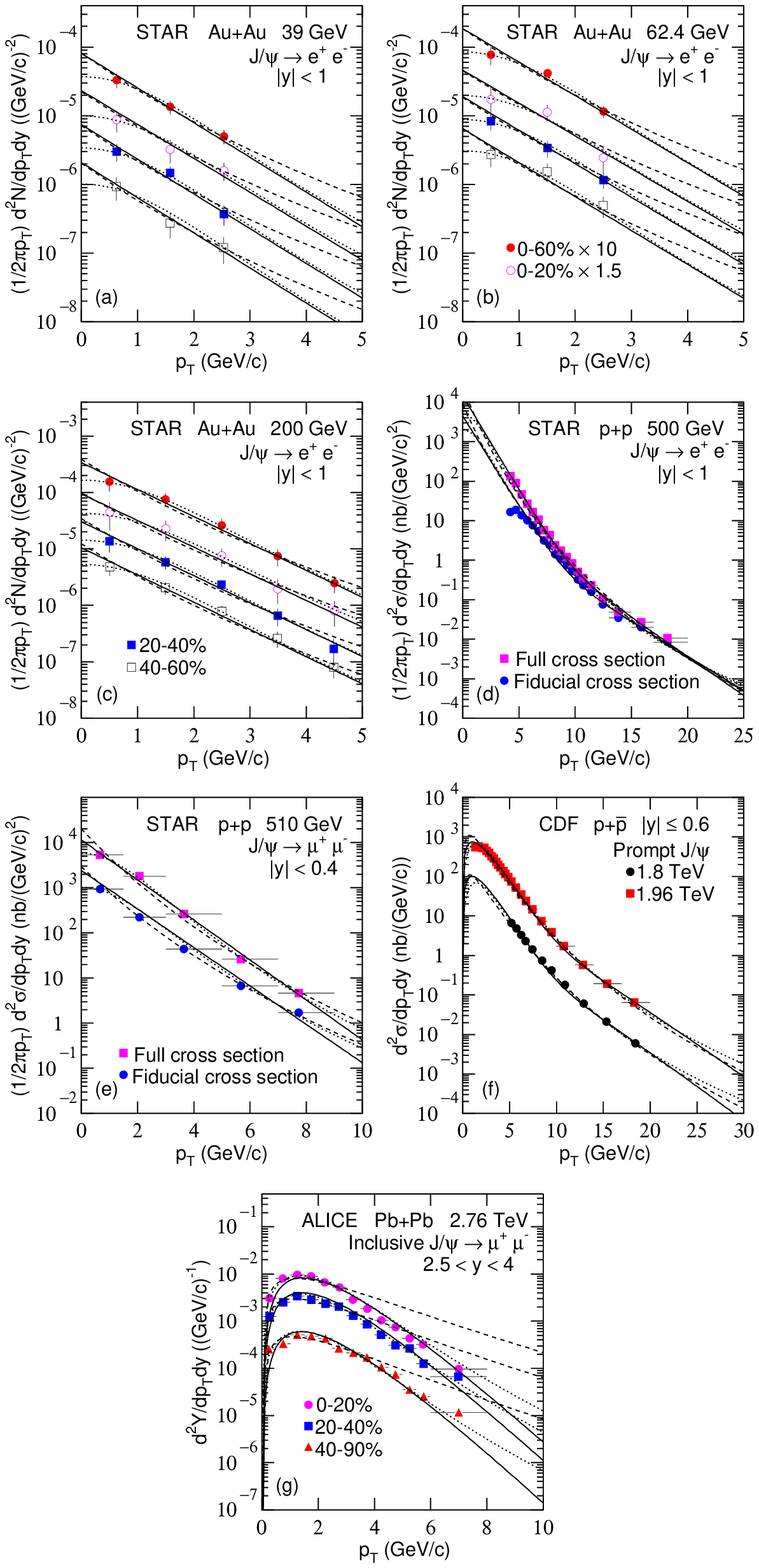}
\end{center}
{\small Fig. 1. Transverse momentum spectra, (a)--(c) $(1/2\pi
p_T)d^2N/dp_Tdy$, (d)(e) $(1/2\pi p_T)d^2\sigma/dp_Tdy$, (f)
$d^2\sigma/dp_Tdy$, and (g) $d^2Y/dp_Tdy$ of $J/\psi$ in different
decay or production modes in (a)--(c) Au+Au, (d)(e) $p$+$p$, (f)
$p$+$\overline{p}$, and (g) Pb+Pb collisions at energies (a) 39,
(b) 62.4, (c) 200, (d) 500, and (e) 510 GeV, as well as (f) 1.8
and 1.96, and (g) 2.76 TeV. The experimental data represented by
the symbols are measured by the (a)--(e) STAR~\cite{27,28}, (f)
CDF~\cite{29a,29b}, and (g) ALICE Collaborations~\cite{30} with
different centrality classes such as (a)--(c) 0-20\%, 20-40\%,
40-60\%, and 0-60\%, and (g) 0-20\%, 20-40\%, and 40-90\%, as well
as with different cross sections e.g. (d)(e) full and fiducial
cross sections. Some data points are scaled by different amounts
marked in the panels. The data points are fitted by the
(two-component) Erlang distribution (Eq. (3), the solid curve),
the Hagedorn function (Eq. (4), the dashed curve), and the
Tsallis-Levy function (Eq. (6), the dotted curve), respectively.}
\end{figure*}

\begin{figure*}[htbp]
\begin{center}
\includegraphics[width=9.5cm]{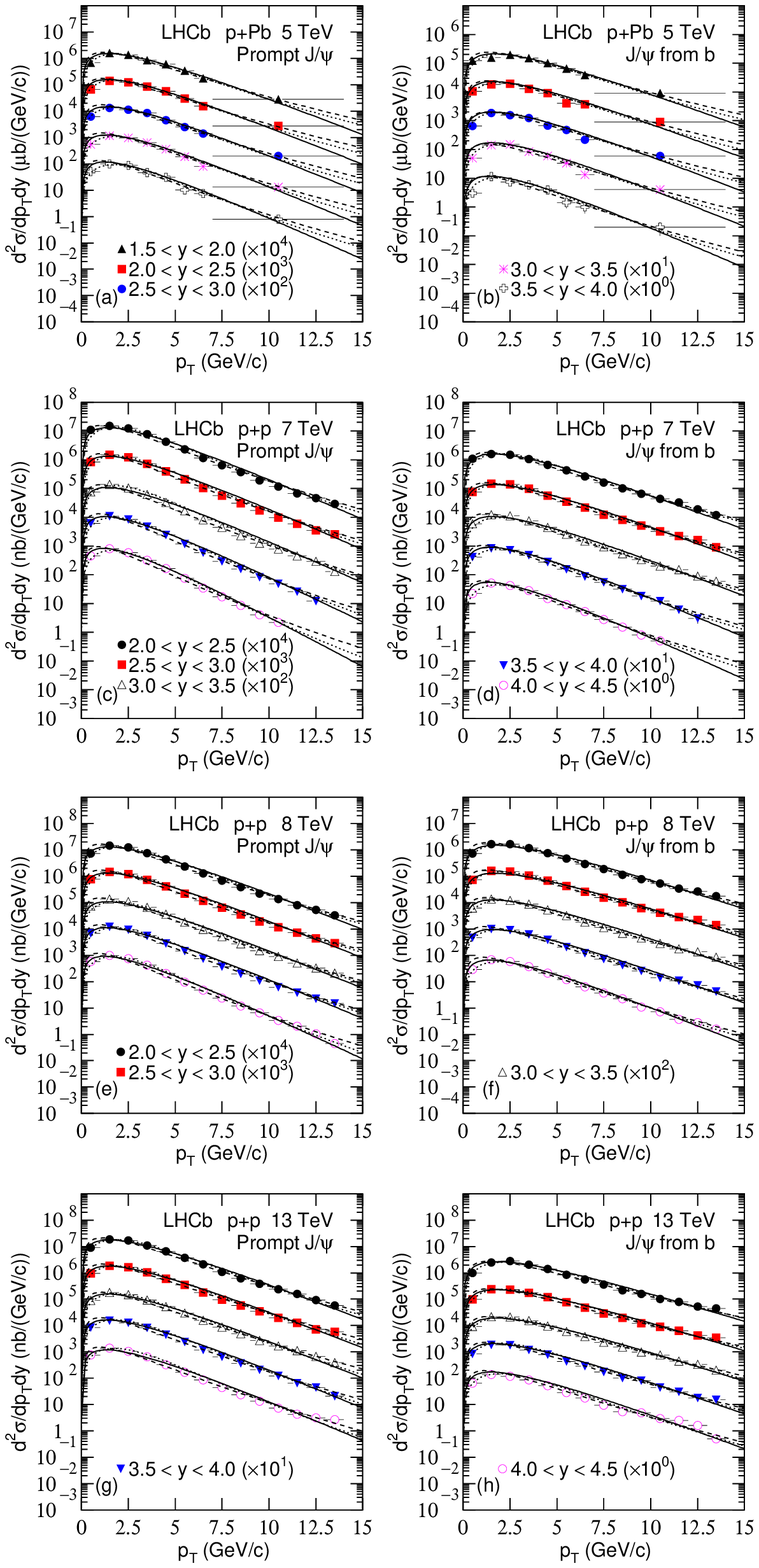}
\end{center}
{\small Fig. 2. Transverse momentum spectra, $d^2\sigma/dp_Tdy$,
of $J/\psi$ in different production modes in (a)(b) $p$+Pb and
(c)--(h) $p$+$p$ collisions at (a)(b) 5, (c)(d) 7, (e)(f) 8, and
(g)(h) 13 TeV. The different symbols represent the experimental
data measured by the LHCb Collaboration~\cite{31,32,33,34} in the
rapidity intervals of (a)(b) $1.5<y<2.0$, $2.0<y<2.5$,
$2.5<y<3.0$, $3.0<y<3.5$, and $3.5<y<4.0$ and (c)--(h)
$2.0<y<2.5$, $2.5<y<3.0$, $3.0<y<3.5$, $3.5<y<4.0$, and
$4.0<y<4.5$, and scaled by different amounts marked in the panels.
The solid, dashed, and dotted curves represent our results fitted
by the Erlang distribution (Eq. (3)), the Hagedorn function (Eq.
(4)), and the Tsallis-Levy function (Eq. (6)), respectively.}
\end{figure*}

\begin{figure*}[htbp]
\begin{center}
\includegraphics[width=9.5cm]{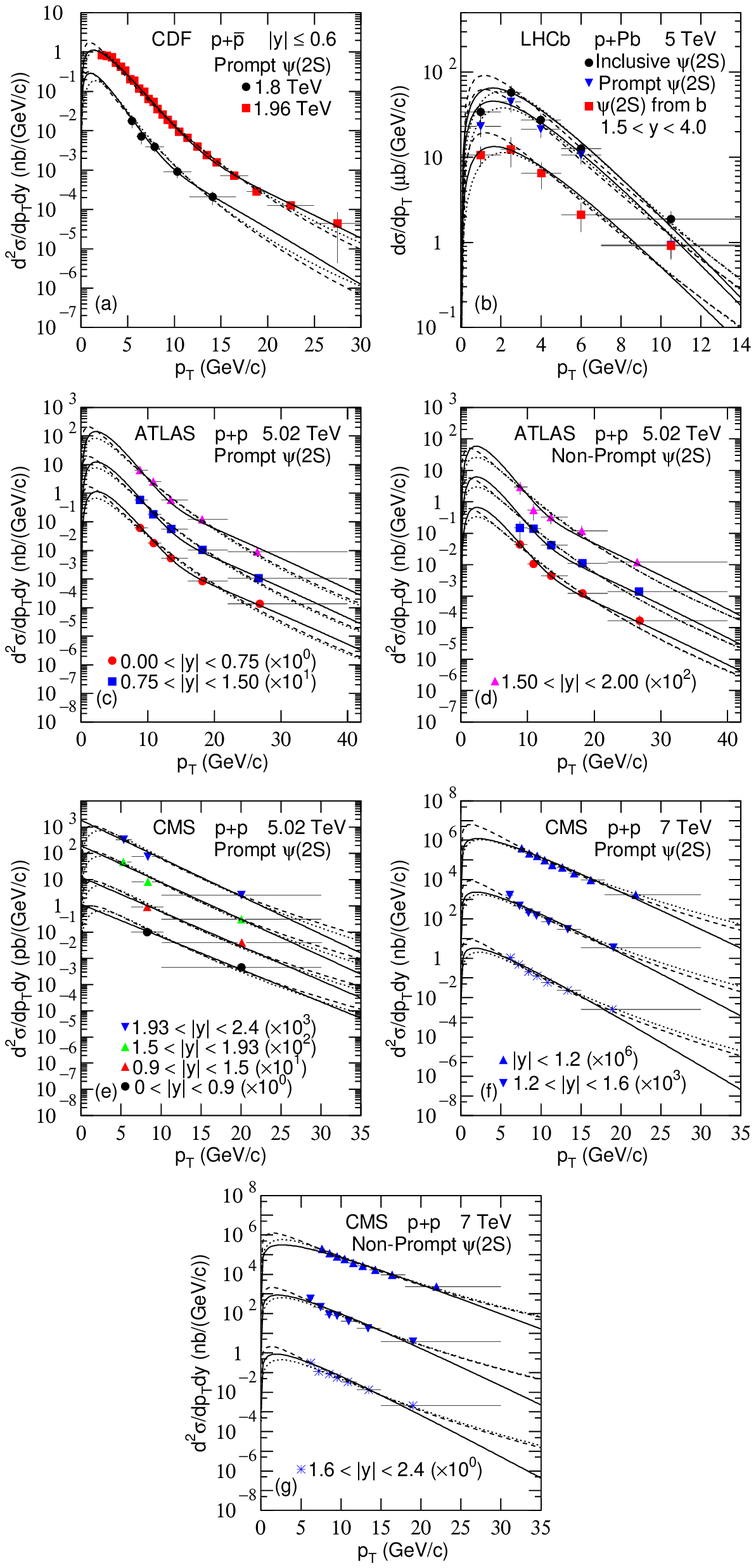}
\end{center}
{\small Fig. 3. Transverse momentum spectra, (a)(c)--(g)
$d^2\sigma/dp_Tdy$ and (b) $d\sigma/dp_T$, of $\psi(2S)$ from
different production modes in (a) $p$+$\overline{p}$, (b) $p$+Pb,
and (c)--(g) $p$+$p$ collisions at energies (a) 1.8 and 1.96, (b)
5, (c)--(e) 5.02, and (f)--(g) 7 TeV. The different symbols
represent the experimental data of the (a)(c)(e)(f) prompt
$\psi(2S)$, (b) inclusive $\psi(2S)$, prompt $\psi(2S)$, and
$\psi(2S)$ from $b$, and (d)(g) non-prompt $\psi(2S)$ measured by
the (a) CDF~\cite{29a,29b}, (b) LHCb~\cite{35}, (c)(d)
ATLAS~\cite{36}, and (e)--(g) CMS Collaborations~\cite{37,38a} in
(a) $p$+$\overline{p}$, (b) $p$+Pb, (c)--(g) $p$+$p$ collisions at
$\sqrt{s_{NN}}$ or $\sqrt{s}=$ (a) 1.8 and 1.96, (b) 5, (c)--(e)
5.02, and (f)(g) 7 TeV with (a) $|y|<0.6$, (b) $1.5<y<4.0$, (c)(d)
$0.00<|y|<0.75$, $0.75<|y|<1.50$, and $1.50<|y|<2.00$, (e)
$0<|y|<0.9$, $0.9<|y|<1.5$, $1.5<|y|<1.93$, and $1.93<|y|<2.4$,
and (f)(g) $|y|<1.2$, $1.2<|y|<1.6$, and $1.6<|y|<2.4$, where
different collaborations have used different precisions for the
rapidity intervals. Some data points are scaled by different
amounts marked in the panels. The solid, dashed, and dotted curves
represent our results fitted by the (two-component) Erlang
distribution (Eq. (3)), the Hagedorn function (Eq. (4)), and the
Tsallis-Levy function (Eq. (6)), respectively.}
\end{figure*}

\begin{figure*}[!htb]
\begin{center}
\includegraphics[width=9.5cm]{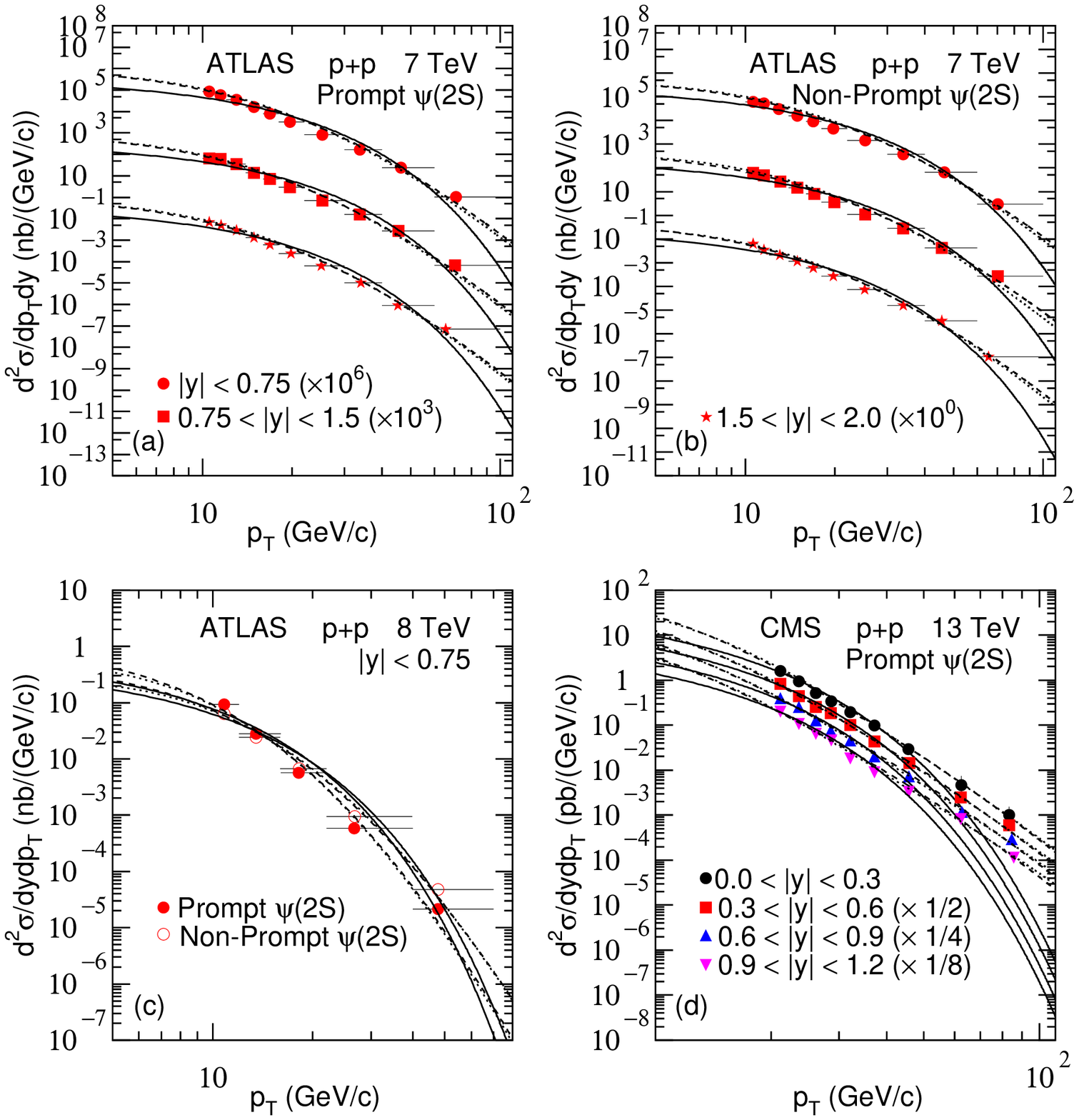}
\end{center}
{\small Fig. 4. Same as Fig. 3, but showing the results at (a)(b)
7, (c) 8, and (d) 13 TeV. The different symbols represent the
experimental data of (a)(d) prompt $\psi(2S)$, (b) non-prompt
$\psi(2S)$, and (c) prompt $\psi(2S)$ and non-prompt $\psi(2S)$
measured by the (a)--(c) ATLAS~\cite{38b,39} and (d) CMS
Collaborations~\cite{40} in (a)(b) $|y|<0.75$, $0.75<|y|<1.5$, and
$1.5<|y|<2.0$, (c) $|y|<0.75$, and (d) $0.0<|y|<0.3$,
$0.3<|y|<0.6$, $0.6<|y|<0.9$, and $0.9<|y|<1.2$, where some data
points are scaled by different amounts marked in the panels. The
data points are fitted by the Erlang distribution (Eq. (3), the
solid curve), the Hagedorn function (Eq. (4), the dashed curves),
and the Tsallis-Levy function (Eq. (6), the dotted curves). In
particular, the two-component Hagedorn function (Eq. (5), the
dashed curves) and the two-component Tsallis-Levy function (Eq.
(7), the dotted curves) are used in Fig. 4(d).}
\end{figure*}

\begin{figure*}[!htb]
\begin{center}
\includegraphics[width=9.5cm]{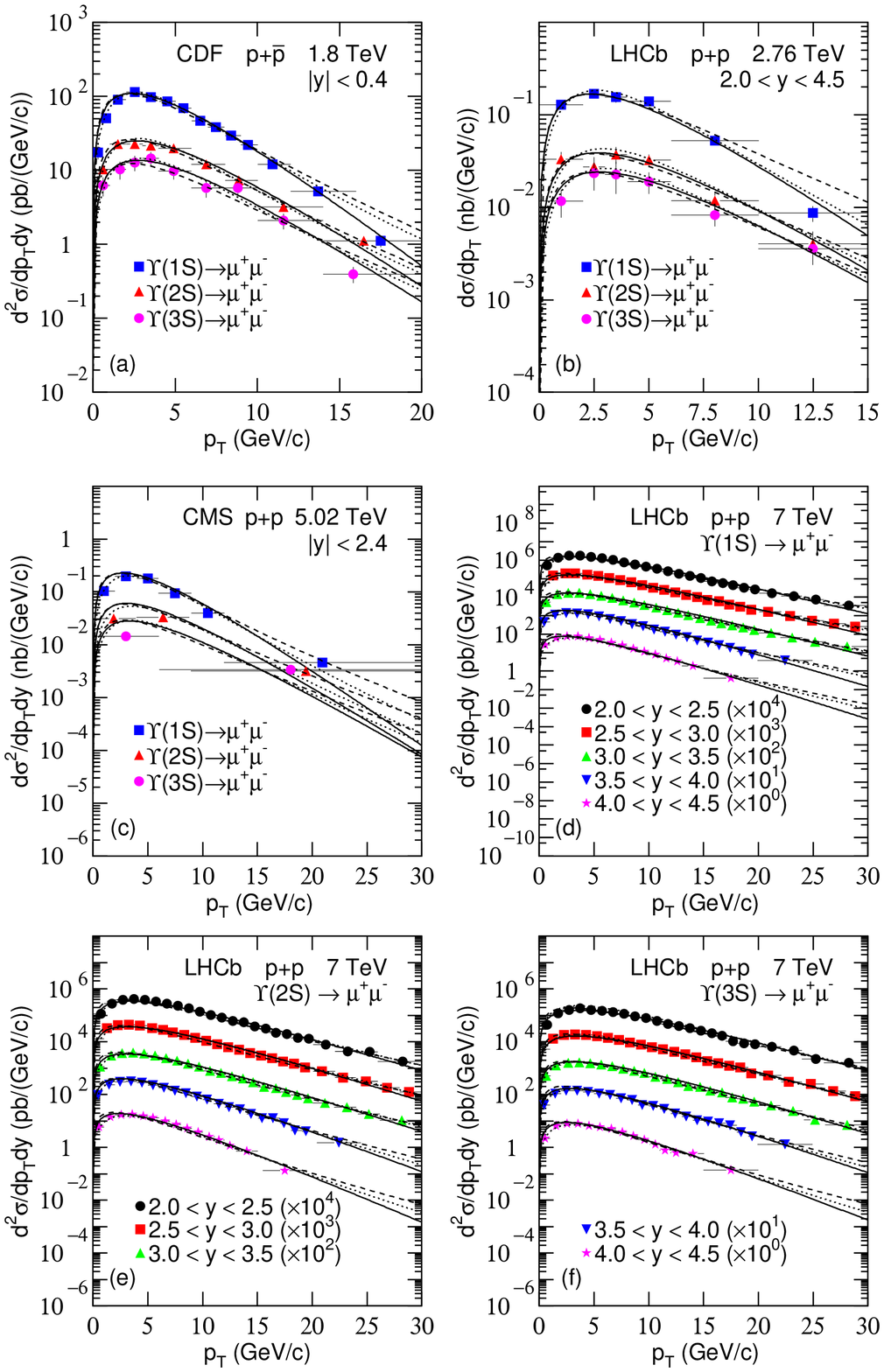}
\end{center}
{\small Fig. 5. Transverse momentum spectra, (a)(c)--(f)
$d^2\sigma/dp_Tdy$ and (b) $d\sigma/dp_T$, of (a)--(c)
$\Upsilon(nS, n=1,2,3)$ $\rightarrow \mu^+\mu^-$, (d)
$\Upsilon(1S)$ $\rightarrow \mu^+\mu^-$, (e) $\Upsilon(2S)$
$\rightarrow \mu^+\mu^-$, and (f) $\Upsilon(3S)$ $\rightarrow
\mu^+\mu^-$ in (a) $p$+$\overline{p}$ and (b)--(f) $p$+$p$
collisions at (a) 1.8, (b) 2.76, (c) 5.02, and (d)--(f) 7 TeV. The
symbols shown in panels (a)--(c) represent the experimental data
measured by the (a) CDF~\cite{41}, (b) LHCb~\cite{42}, and (c) CMS
Collaborations~\cite{43} in (a) $p$+$\overline{p}$ and (b)(c)
$p$+$p$ collisions in (a) $|y|<0.4$, (b) $2.0<y<4.5$, and (c)
$|y|<2.4$ respectively. The symbols shown in panels (d)--(f)
represent the experimental data measured by the LHCb
Collaboration~\cite{44} in $p$+$p$ collisions in $2.0<y<2.5$,
$2.5<y<3.0$, $3.0<y<3.5$, $3.5<y<4.0$, and $4.0<y<4.5$ and scaled
by different amounts shown in the panels. The data points are
fitted by the Erlang distribution (Eq. (3)), the Hagedorn function
(Eq. (4)), and the Tsallis-Levy function (Eq. (6)) by the solid,
dashed, and dotted curves, respectively.}
\end{figure*}

\begin{figure*}[htbp]
\begin{center}
\includegraphics[width=9.5cm]{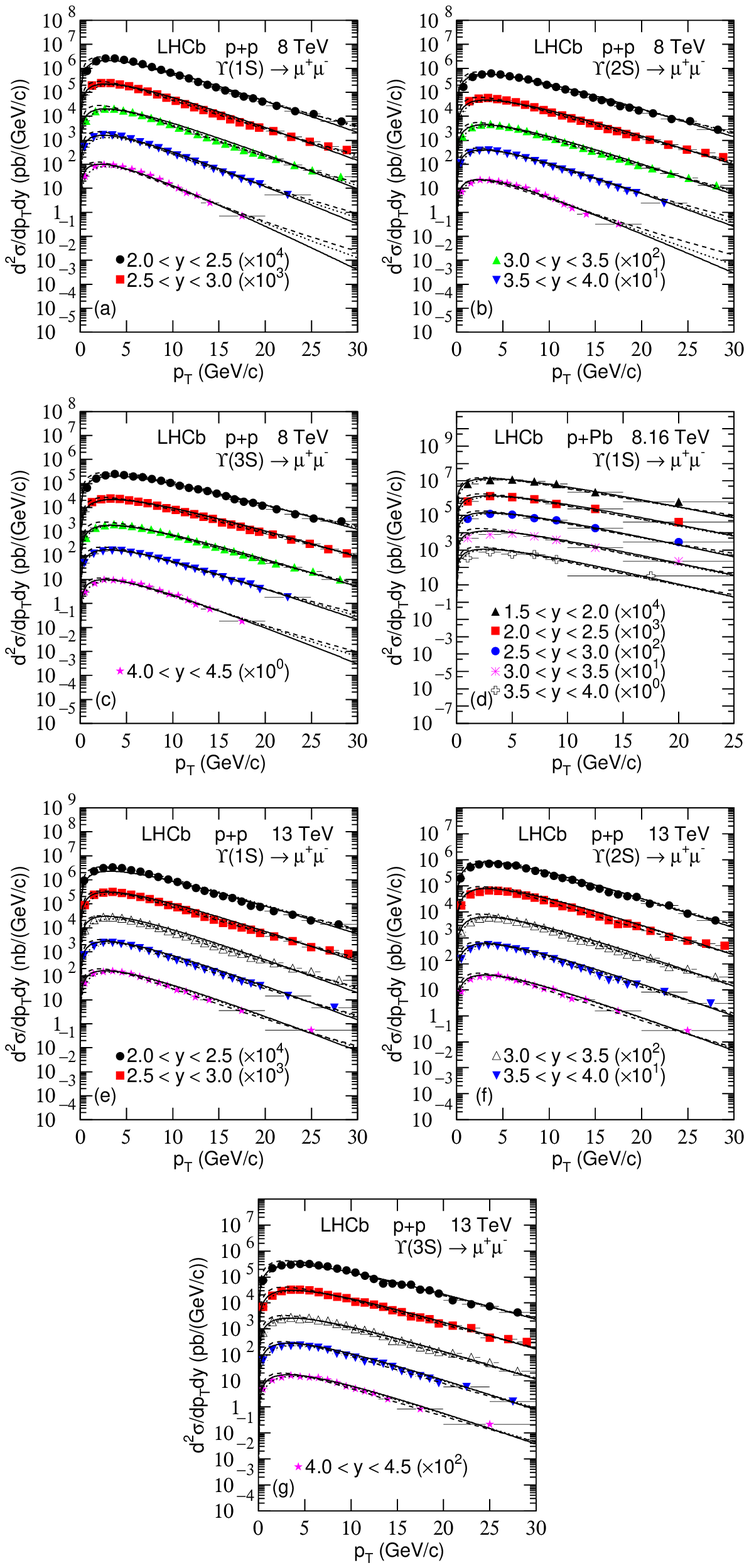}
\end{center}
{\small Fig. 6. Transverse momentum spectra, $d^2\sigma/dp_Tdy$,
of (a)(d)(e) $\Upsilon(1S)$ $\rightarrow \mu^+\mu^-$, (b)(f)
$\Upsilon(2S)$ $\rightarrow \mu^+\mu^-$, and (c)(g) $\Upsilon(3S)$
$\rightarrow \mu^+\mu^-$ in (a)--(c) and (e)--(g) $p$+$p$ and (d)
$p$+Pb collisions at (a)--(c) 8, (d) 8.16, and (e)--(g) 13 TeV.
The symbols represent the experimental data measured by the LHCb
Collaboration~\cite{44,45,46}. The rapidity intervals for panels
(a)--(c) and (e)--(f) are $2.0<y<2.5$, $2.5<y<3.0$, $3.0<y<3.5$,
$3.5<y<4.0$, and $4.0< y<4.5$. The rapidity intervals for panel
(d) are $1.5< y<2.0$ $2.0<y<2.5$, $2.5<y<3.0$, $3.0<y<3.5$, and
$3.5<y<4.0$. Different sets of data points are scaled by different
amounts shown in the panels. The data points are fitted by the
Erlang distribution (Eq. (3)), the Hagedorn function (Eq. (4)),
and the Tsallis-Levy function (Eq. (6)) by the solid, dashed, and
dotted curves, respectively.}
\end{figure*}

\begin{figure*}[!htb]
\begin{center}
\includegraphics[width=9.5cm]{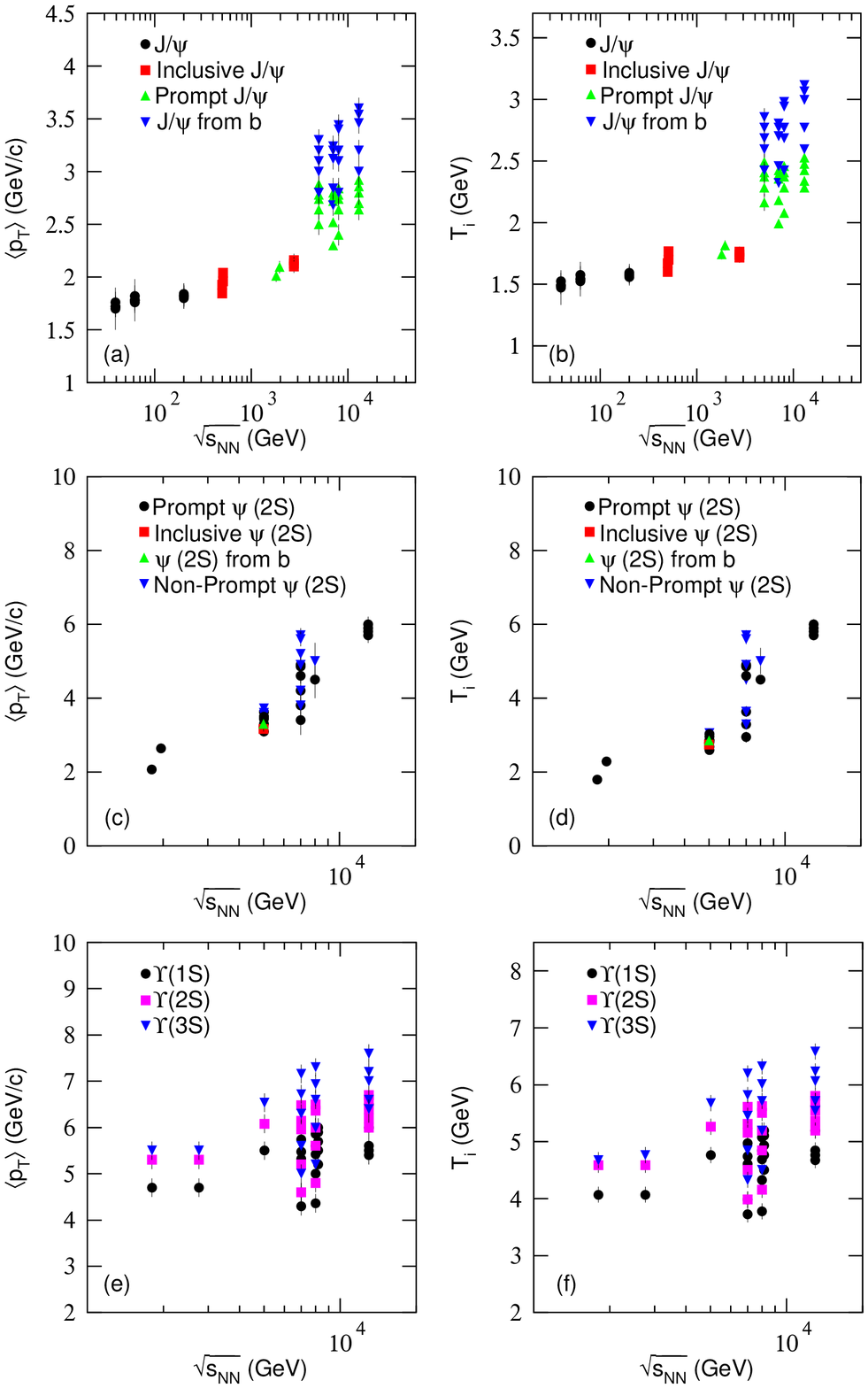}
\end{center}
{\small Fig. 7. Dependences of (a)(c)(e) $\langle p_T\rangle$ and
(b)(d)(f) $T_i$ on $\sqrt{s_{NN}}$ (or $\sqrt{s}$) for (a)(b)
$J/\psi$, (c)(d) $\psi(2S)$, and (e)(f) $\Upsilon(nS, n=1,2,3)$.
The different symbols represent the parameter values derived from
Figs. 1--6 and listed in Tables A1 and A2 where only the
(two-component) Erlang distribution in the $p_T$ region of data
available is used as an example. By using the mentioned three
functions which fit the data good enough in the $p_T$ region of
data available, one can obtain similar $\langle p_T\rangle$ (or
$T_i$) within a systematic uncertainty of 8\%.}
\end{figure*}

\begin{figure*}[!htb]
\begin{center}
\includegraphics[width=9.5cm]{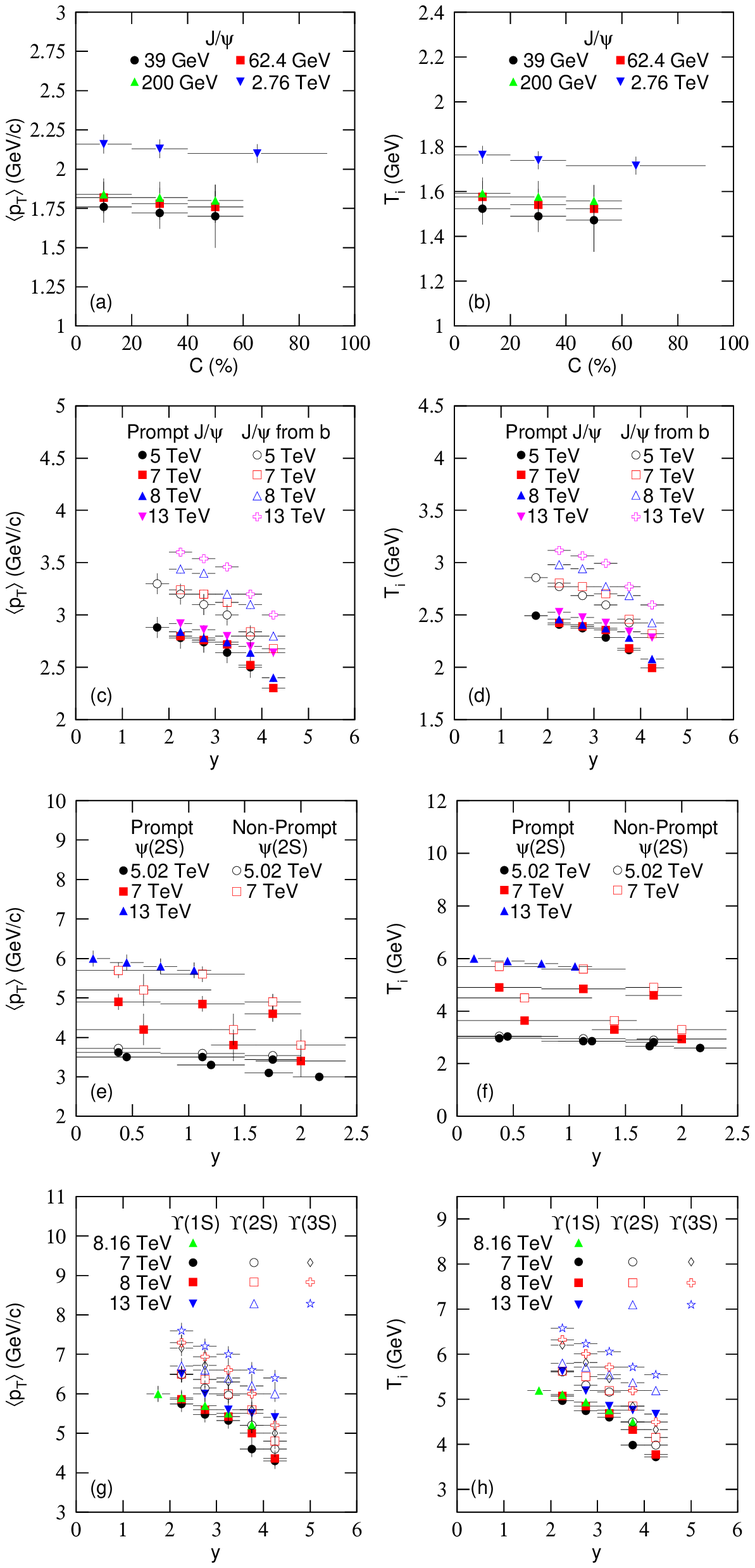}
\end{center}
{\small Fig. 8. Same as Fig. 7, but showing the dependences of
(a)(c)(e)(g) $\langle p_T\rangle$ and (b)(d)(f)(h) $T_i$ on (a)(b)
$C$ and (c)--(h) $y$ for (a)--(d) $J/\psi$, (e)(f) $\psi(2S)$, and
(g)(h) $\Upsilon(nS, n=1,2,3)$.}
\end{figure*}

The $p_T$ spectra, $d^2\sigma/dp_Tdy$, of (a)(c)(e)(g) prompt
$J/\psi$ and (b)(d)(f)(h) $J/\psi$ from $b$ produced in (a)(b)
$p$+Pb and (c)--(h) $p$+$p$ collisions at $\sqrt{s_{NN}}$ or
$\sqrt{s}=$ (a)(b) 5, (c)(d) 7, (e)(f) 8, and (g)(h) 13 TeV are
presented in Fig. 2. The symbols represent the experimental
data~\cite{31,32,33,34} and the curve are our fitted results. The
method of least square is used to obtain the best parameter values
which are listed in Tables A1 and A2 with $\chi^2$ and ndof. One
can see that the experimental $p_T$ spectra of $J/\psi$ via
different production modes in different rapidity intervals in
$p$+$p$ and $p$+Pb collisions at high energies are approximately
fitted by the Erlang distribution, the Hagedorn function, and the
Tsallis-Levy function.

Figure 3 shows the $p_T$ spectra, (a)(c)--(g) $d^2\sigma/dp_Tdy$
and (b) $d\sigma/dp_T$, of $\psi(2S)$ via different production
modes. The symbols represent the experimental
data~\cite{29a,29b,35,36,37,38a} and the curve are our fitted
results. The values of free parameters are listed in Tables A1 and
A2 with $\chi^2$ and ndof. One can see that the experimental $p_T$
spectra of $\psi(2S)$ via different production modes in different
rapidity intervals in $p$+$p$, $p$+$\overline{p}$, and $p$+Pb
collisions at high energies are approximately fitted by the
(two-component) Erlang distribution, the Hagedorn function, and
the Tsallis-Levy function.

Figure 4 shows the $p_T$ spectra, $d^2\sigma/dp_Tdy$, of
$\psi(2S)$ via different production modes in $p$+$p$ collisions at
$\sqrt{s}=$ (a)(b) 7, (c) 8, and (d) 13 TeV. The symbols represent
the experimental data~\cite{38b,39,40} and the curve are our
fitted results. The values of free parameters are listed in Tables
A1 and A2 with $\chi^2$ and ndof. For Fig. 4(d), the two-component
Eqs. (5) and (7) are used, where the free parameters for the first
(second) component are listed in the first (second) row. One can
see that the experimental $p_T$ spectra of $\psi(2S)$ via
different production modes in different rapidity intervals in
$p$+$p$ collisions at high energies are also approximately fitted
by the Erlang distribution, the (two-component) Hagedorn function,
and the (two-component) Tsallis-Levy function.

In Fig. 5, the $p_T$ spectra, (a)(c)--(f) $d^2\sigma/dp_Tdy$ and
(b) $d\sigma/dp_T$, of (a)--(c) $\Upsilon(nS, n=1,2,3)$
$\rightarrow \mu^+\mu^-$, (d) $\Upsilon(1S)$ $\rightarrow
\mu^+\mu^-$, (e) $\Upsilon(2S)$ $\rightarrow \mu^+\mu^-$, and (f)
$\Upsilon(3S)$ $\rightarrow \mu^+\mu^-$ induced in (a)
$p$+$\overline{p}$ and (b)--(f) $p$+$p$ collisions at $\sqrt{s}=$
(a) 1.8, (b) 2.76, (c) 5.02, and (d)--(f) 7 TeV are given. The
symbols represent the experimental data~\cite{41,42,44,43} and the
curve are our fitted results. The parameter values are listed in
Tables A1 and A2 with $\chi^2$ and ndof. One can see that the
experimental $p_T$ spectra of $\Upsilon(nS, n=1,2,3)$ $\rightarrow
\mu^+\mu^-$ in different rapidity intervals in $p$+$p$ and
$p$+$\overline{p}$ collisions at high energies are approximately
fitted by the Erlang distribution, the Hagedorn function, and the
Tsallis-Levy function.

In Fig. 6, the $p_T$ spectra, $d^2\sigma dp_Tdy$, of (a)(d)(e)
$\Upsilon(1S)$ $\rightarrow$ $\mu^+\mu^-$, (b)(f) $\Upsilon(2S)$
$\rightarrow$ $\mu^+\mu^-$, and (c)(g) $\Upsilon(3S)$
$\rightarrow$ $\mu^+\mu^-$ induced in (a)--(c) and (e)--(g)
$p$+$p$ and (d) $p$+Pb collisions at $\sqrt{s_{NN}}$ or
$\sqrt{s}=$ (a)--(c) 8, (d) 8.16, and (e)--(g) 13 TeV are given.
The symbols represent the experimental data~\cite{44,45,46} and
the curve are our fitted results. The parameter values are listed
in Table A1 and A2 with $\chi^2$ and ndof. Once again, one can see
that the experimental $p_T$ spectra of $\Upsilon(nS, n=1,2,3)$
$\rightarrow \mu^+\mu^-$ in different rapidity intervals in
$p$+$p$ and $p$+Pb collisions at high energies are approximately
fitted by the Erlang distribution, the Hagedorn function, and the
Tsallis-Levy function.

Before discussing the trends of parameters, we would like to point
out the usability of the concept of temperature in $p$+$p$
($p$+$\overline{p}$) collisions which are small in size. As in
refs.~\cite{55a,55b,55c,55d}, in this work, we have treated
$p$+$p$ ($p$+$\overline{p}$) collisions as where a medium was
formed, or at least there is some degree of thermalization, enough
to have a temperature for the emission source. On the other hand,
the temperature parameter of the emission source is a reflection
of average kinetic energy of given particles. This means that we
may use the concept of temperature. Even if the collision system
is not enough large, we may use the temperature parameter to
characterize the average kinetic energy of given particles over
many events.

Figure 7 shows the dependences of (a)(c)(e) $\langle p_T\rangle$
and (b)(d)(f) $T_i$ on $\sqrt{s_{NN}}$ (or $\sqrt{s}$) for (a)(b)
$J/\psi$, (c)(d) $\psi(2S)$, and (e)(f) $\Upsilon(nS, n=1,2,3)$.
The different symbols represent the parameter values derived from
free parameters extracted from Figs. 1--6 and listed in Tables A1
and A2, where only the (two-component) Erlang distribution in the
$p_T$ region of data available is used as an example. It is
expected that the results corresponding to the Hagedorn and
Tsallis-Levy functions are very close to the plot, because the two
functions also describe approximately the data. As what we
discussed in subsection 2.5, no matter what functions are used to
fit the experimental data, one should obtain similar $\langle
p_T\rangle$ (or $T_i$), if different functions fit the data good
enough in the $p_T$ region of data available. By using the
mentioned three functions which fit the data good enough, one can
obtain $\langle p_T\rangle$ (or $T_i$) within a systematic
uncertainty of 8\%. One can see from Figure 7 that $\langle
p_T\rangle$ and $T_i$ increases significantly with the increase of
collision energy. Meanwhile, $\langle p_T\rangle$ and $T_i$
increases with the increase of particle mass.

Figure 8 is the same as Fig. 7, but showing the dependences of
(a)(c)(e)(g) $\langle p_T\rangle$ and (b)(d)(f)(h) $T_i$ on (a)(b)
centrality $C$ and (c)--(h) rapidity $y$ for (a)--(d) $J/\psi$,
(e)(f) $\psi(2S)$, and (g)(h) $\Upsilon(nS, n=1,2,3)$. The
different symbols represent the parameter values derived from free
parameters extracted from Figs. 1--6 and listed in Tables A1 and
A2, where only the the (two-component) Erlang distribution is used
as an example. One can see that $\langle p_T\rangle$ and $T_i$
increases slightly with the increase of event centrality from
peripheral to central collisions, and decreases with the increase
of rapidity from mid-rapidity to forward rapidity. Meanwhile,
$\langle p_T\rangle$ and $T_i$ increases with the increases of
collision energy and particle mass.

The above parameter tendencies show that the temperature is mass
dependent. This is also a reflection of the formation time
dependence. According to the hydrodynamic behavior, ``massive
particles coming out of the system earlier in time with smaller
radial flow velocities"~\cite{55d}. This means that with the
increase of mass, the formation time decreases, the temperature
increases, and the flow velocity decreases. It should be noted
that the fact that the massive particles coming out the system
earlier is not caused by the high excitation of the system, but
the leaver over due to the inertia of massive particles, in the
hydrodynamic evolution.

We may explain the tendency of derived $\langle p_T\rangle$ and
$T_i$ which have similar tendency with $p_T$. With the increase of
collision energy, the violent degree of collisions increases
significantly due to large energy transfer, which results in the
obvious increase of $p_T$. With the increase of centrality, the
degree of multiple-scattering increases due to more participant
nucleons and produced particles taking part in the scattering
process, which results in slight increase of emission angle and
then slight increase of $p_T$. With the increase of rapidity, the
energy transfer decreases due to larger penetrability between
participant nucleons. Meanwhile, the degree of multiple-scattering
also decreases due to less produced particles taking part in the
scattering process. These two factors result in the decrease of
$p_T$. It is natural that $p_T$ increases with the increase of
$m_0$.

The free parameters in the Erlang distribution are directly
reflected in $\langle p_T\rangle$, which will not be discussed
anymore. The free parameters in the Hagedorn and Tsallis-Levy
functions will be discussed in the Appendix 2, because no clearly
physical conclusion can be drawn from them at present, though the
tendencies of them can be seen from this work.
\\

{\section{Summary and conclusions}}

In summary, the transverse momentum spectra of $J/\psi$,
$\psi(2S)$, and $\Upsilon(nS, n=1,2,3)$ produced in $p$+$p$,
$p$+$\overline{p}$, $p$+Pb, Au+Au, and Pb+Pb collisions over an
energy range from 39 GeV to 13 TeV have been analyzed by the
(two-component) Erlang distribution, the Hagedorn function, and
the Tsallis-Levy function. The function results are approximately
in agreement with the experimental data measured by several
international collaborations. The values of related parameters are
extracted from the fits and the excitation functions of these
parameters are obtained.

The excitation functions of parameters $\langle p_T\rangle$ and
$T_i$ increase from 39 GeV to 13 TeV. Meanwhile, $\langle
p_T\rangle$ and $T_i$ increase (slightly) with event centrality
and particle mass and decrease from mid-rapidity to forward
rapidity. These tendencies render that these parameters describe
the excitation and expansion degrees of the system. At higher
energy, larger energy transfer had happened, which results in
higher excitation and expansion degrees of the system. In central
collisions and at mid-rapidity, larger energy transfer and further
multiple-scattering had happened, which also results in higher
excitation and expansion degrees of the system.

The parameters $p_0$ ($T$) and $n_0$ ($n$) increase with the
collision energy, which reflects the degree of energy deposition
and transfer. In given collisions, there is a negative correlation
between $p_0$ ($T$) and $n_0$ ($n$). At different energies, there
is a positive correlation between $p_0$ ($T$) and $n_0$ ($n$).
Indeed, there are correlations between $p_0$ ($T$) and $n_0$ ($n$)
when we determine these parameters. The correlation between $p_0$
($T$) and $n_0$ ($n$) is similar to that between kinetic
freeze-out temperature and transverse flow velocity. If $p_0$
($T$) is similar to kinetic freeze-out temperature, $n_0$ ($n$)
should be similar to transverse flow velocity.
\\
\\
{\bf Data Availability}

The data used to support the findings of this study are included
within the article and are cited at relevant places within the
text as references.
\\
\\
{\bf Ethical Approval}

The authors declare that they are in compliance with ethical
standards regarding the content of this paper.
\\
\\
{\bf Disclosure}

The funding agencies have no role in the design of the study; in
the collection, analysis, or interpretation of the data; in the
writing of the manuscript; or in the decision to publish the
results.
\\
\\
{\bf Conflicts of Interest}

The authors declare that there are no conflicts of interest
regarding the publication of this paper.
\\
\\
{\bf Acknowledgments}

This work was supported by the National Natural Science Foundation
of China under Grant Nos. 11575103 and 11947418, the Scientific
and Technological Innovation Programs of Higher Education
Institutions in Shanxi (STIP) under Grant No. 201802017, the
Shanxi Provincial Natural Science Foundation under Grant No.
201901D111043, and the Fund for Shanxi ``1331 Project" Key
Subjects Construction.
\\
\\

\end{multicols}

\clearpage


{\bf Appendix 1: Tables of parameters.}
\\

{\scriptsize Table A1. Values of $\langle p_t\rangle_1$, $\langle
p_t\rangle_2$, $N_1$, $k_E$, $\langle p_T\rangle$, $T_i$, and
$\chi^2$/ndof corresponding to the solid curves in Figs. 1 and 2.
In all cases, $N_2=2$ which is not listed in the table. In the
case of ndof $\leq0$, we use ``$-$" to mention.} \vspace{-4mm}
\begin{center}
{\scriptsize
\begin{tabular} {ccccccccc}\\ \hline\hline Figure & Main selection & $\langle p_t\rangle_1$ (GeV/$c$) & $\langle p_t\rangle_2$ (GeV/$c$) & $N_1$ & $k_E$ & $\langle p_T\rangle$ (GeV/$c$) & $T_i$ (GeV) & $\chi^2$/ndof\\
\hline
1(a) & 0--20\%  & $0.88\pm0.05$ & $-$ & $2\pm0$ & $1$ & $1.760\pm0.100$ & $1.520\pm0.070$ & $1.87/-$\\
     & 20--40\% & $0.86\pm0.05$ & $-$ & $2\pm0$ & $1$ & $1.720\pm0.100$ & $1.490\pm0.070$ & $1.67/-$\\
     & 40--60\% & $0.85\pm0.10$ & $-$ & $2\pm0$ & $1$ & $1.700\pm0.200$ & $1.472\pm0.140$ & $0.38/-$\\
     & 0--60\%  & $0.86\pm0.05$ & $-$ & $2\pm0$ & $1$ & $1.720\pm0.100$ & $1.490\pm0.070$ & $1.34/-$\\
1(b) & 0--20\%  & $0.91\pm0.05$ & $-$ & $2\pm0$ & $1$ & $1.820\pm0.100$ & $1.576\pm0.070$ & $2.10/-$\\
     & 20--40\% & $0.89\pm0.05$ & $-$ & $2\pm0$ & $1$ & $1.780\pm0.100$ & $1.542\pm0.070$ & $1.05/-$\\
     & 40--60\% & $0.88\pm0.05$ & $-$ & $2\pm0$ & $1$ & $1.760\pm0.100$ & $1.524\pm0.070$ & $2.64/-$\\
     & 0--60\%  & $0.89\pm0.01$ & $-$ & $2\pm0$ & $1$ & $1.780\pm0.200$ & $1.542\pm0.140$ & $2.81/-$\\
1(c) & 0--20\%  & $0.92\pm0.05$ & $-$ & $2\pm0$ & $1$ & $1.840\pm0.100$ & $1.593\pm0.070$ & $0.66/2$\\
     & 20--40\% & $0.91\pm0.05$ & $-$ & $2\pm0$ & $1$ & $1.820\pm0.100$ & $1.576\pm0.070$ & $2.47/2$\\
     & 40--60\% & $0.90\pm0.05$ & $-$ & $2\pm0$ & $1$ & $1.800\pm0.100$ & $1.559\pm0.070$ & $2.65/2$\\
     & 0--60\%  & $0.91\pm0.05$ & $-$ & $2\pm0$ & $1$ & $1.820\pm0.100$ & $1.576\pm0.070$ & $1.99/2$\\
1(d) & Full cross section     & $0.91\pm0.02$ & $2.20\pm0.05$ & $2\pm0$ & $0.99\pm0.01$ & $1.846\pm0.034$ & $1.663\pm0.037$ & $13.40/14$\\
     & Fiducial cross section & $0.95\pm0.01$ & $2.50\pm0.05$ & $2\pm0$ & $0.99\pm0.01$ & $1.931\pm0.030$ & $1.709\pm0.040$ & $54.91/14$\\
1(e) & Full cross section     & $0.98\pm0.02$ & $-$ & $2\pm0$ & $1$ & $1.960\pm0.040$ & $1.697\pm0.030$ & $5.50/2$\\
     & Fiducial cross section & $1.02\pm0.02$ & $-$ & $2\pm0$ & $1$ & $2.040\pm0.040$ & $1.767\pm0.030$ & $14.50/2$\\
1(f) & 1.8 TeV  & $0.98\pm0.05$ & $2.30\pm0.10$ & $2\pm0$ & $0.98\pm0.01$ & $2.013\pm0.056$ & $1.743\pm0.040$ & $6.54/6$\\
     & 1.96 TeV & $1.02\pm0.05$ & $2.41\pm0.10$ & $2\pm0$ & $0.98\pm0.01$ & $2.096\pm0.057$ & $1.815\pm0.041$ & $17.34/19$\\
1(g) & 0--20\%  & $0.72\pm0.02$ & $-$ & $3\pm0$ & $1$ & $2.160\pm0.060$ & $1.764\pm0.040$ & $25.90/10$\\
     & 20--40\% & $0.71\pm0.02$ & $-$ & $3\pm0$ & $1$ & $2.130\pm0.060$ & $1.739\pm0.040$ & $20.90/10$\\
     & 40--90\% & $0.70\pm0.02$ & $-$ & $3\pm0$ & $1$ & $2.100\pm0.060$ & $1.715\pm0.040$ & $30.75/10$\\
\hline
2(a) & $1.5<y<2.0$ & $1.44\pm0.05$ & $-$ & $2\pm0$ & $1$ & $2.880\pm0.100$ & $2.494\pm0.035$ & $4.00/5$\\
     & $2.0<y<2.5$ & $1.39\pm0.05$ & $-$ & $2\pm0$ & $1$ & $2.780\pm0.100$ & $2.408\pm0.035$ & $4.40/5$\\
     & $2.5<y<3.0$ & $1.37\pm0.05$ & $-$ & $2\pm0$ & $1$ & $2.740\pm0.100$ & $2.373\pm0.035$ & $3.95/5$\\
     & $3.0<y<3.5$ & $1.32\pm0.05$ & $-$ & $2\pm0$ & $1$ & $2.640\pm0.100$ & $2.286\pm0.035$ & $3.96/5$\\
     & $3.5<y<4.0$ & $1.25\pm0.05$ & $-$ & $2\pm0$ & $1$ & $2.500\pm0.100$ & $2.165\pm0.035$ & $4.84/5$\\
2(b) & $1.5<y<2.0$ & $1.65\pm0.05$ & $-$ & $2\pm0$ & $1$ & $3.300\pm0.100$ & $2.858\pm0.035$ & $3.01/5$\\
     & $2.0<y<2.5$ & $1.60\pm0.05$ & $-$ & $2\pm0$ & $1$ & $3.200\pm0.100$ & $2.771\pm0.035$ & $5.99/5$\\
     & $2.5<y<3.0$ & $1.55\pm0.05$ & $-$ & $2\pm0$ & $1$ & $3.100\pm0.100$ & $2.685\pm0.035$ & $8.89/5$\\
     & $3.0<y<3.5$ & $1.50\pm0.05$ & $-$ & $2\pm0$ & $1$ & $3.000\pm0.100$ & $2.598\pm0.035$ & $11.77/5$\\
     & $3.5<y<4.0$ & $1.40\pm0.05$ & $-$ & $2\pm0$ & $1$ & $2.800\pm0.100$ & $2.425\pm0.035$ & $13.44/5$\\
2(c) & $2.0<y<2.5$ & $1.40\pm0.02$ & $-$ & $2\pm0$ & $1$ & $2.800\pm0.040$ & $2.425\pm0.014$ & $10.37/11$\\
     & $2.5<y<3.0$ & $1.38\pm0.02$ & $-$ & $2\pm0$ & $1$ & $2.760\pm0.040$ & $2.390\pm0.014$ & $10.43/11$\\
     & $3.0<y<3.5$ & $1.36\pm0.02$ & $-$ & $2\pm0$ & $1$ & $2.720\pm0.040$ & $2.356\pm0.014$ & $10.12/11$\\
     & $3.5<y<4.0$ & $1.26\pm0.02$ & $-$ & $2\pm0$ & $1$ & $2.520\pm0.040$ & $2.182\pm0.014$ & $7.96/10$\\
     & $4.0<y<4.5$ & $1.15\pm0.02$ & $-$ & $2\pm0$ & $1$ & $2.300\pm0.040$ & $1.992\pm0.014$ & $5.97/8$\\
2(d) & $2.0<y<2.5$ & $1.62\pm0.02$ & $-$ & $2\pm0$ & $1$ & $3.240\pm0.040$ & $2.806\pm0.014$ & $7.49/11$\\
     & $2.5<y<3.0$ & $1.60\pm0.02$ & $-$ & $2\pm0$ & $1$ & $3.200\pm0.040$ & $2.771\pm0.014$ & $7.18/11$\\
     & $3.0<y<3.5$ & $1.56\pm0.02$ & $-$ & $2\pm0$ & $1$ & $3.120\pm0.040$ & $2.702\pm0.014$ & $8.82/11$\\
     & $3.5<y<4.0$ & $1.42\pm0.02$ & $-$ & $2\pm0$ & $1$ & $2.840\pm0.040$ & $2.460\pm0.014$ & $5.13/10$\\
     & $4.0<y<4.5$ & $1.34\pm0.02$ & $-$ & $2\pm0$ & $1$ & $2.680\pm0.040$ & $2.321\pm0.014$ & $6.36/8$\\
2(e) & $2.0<y<2.5$ & $1.42\pm0.02$ & $-$ & $2\pm0$ & $1$ & $2.840\pm0.040$ & $2.460\pm0.014$ & $9.57/11$\\
     & $2.5<y<3.0$ & $1.39\pm0.02$ & $-$ & $2\pm0$ & $1$ & $2.780\pm0.040$ & $2.408\pm0.014$ & $7.10/11$\\
     & $3.0<y<3.5$ & $1.37\pm0.02$ & $-$ & $2\pm0$ & $1$ & $2.740\pm0.040$ & $2.373\pm0.014$ & $9.77/11$\\
     & $3.5<y<4.0$ & $1.32\pm0.02$ & $-$ & $2\pm0$ & $1$ & $2.640\pm0.040$ & $2.286\pm0.014$ & $9.62/11$\\
     & $4.0<y<4.5$ & $1.20\pm0.02$ & $-$ & $2\pm0$ & $1$ & $2.400\pm0.040$ & $2.078\pm0.014$ & $7.60/11$\\
2(f) & $2.0<y<2.5$ & $1.72\pm0.02$ & $-$ & $2\pm0$ & $1$ & $3.440\pm0.040$ & $2.979\pm0.014$ & $8.11/11$\\
     & $2.5<y<3.0$ & $1.70\pm0.02$ & $-$ & $2\pm0$ & $1$ & $3.400\pm0.040$ & $2.944\pm0.014$ & $8.99/11$\\
     & $3.0<y<3.5$ & $1.60\pm0.02$ & $-$ & $2\pm0$ & $1$ & $3.200\pm0.040$ & $2.771\pm0.014$ & $9.57/11$\\
     & $3.5<y<4.0$ & $1.55\pm0.02$ & $-$ & $2\pm0$ & $1$ & $3.100\pm0.040$ & $2.685\pm0.014$ & $9.89/11$\\
     & $4.0<y<4.5$ & $1.40\pm0.02$ & $-$ & $2\pm0$ & $1$ & $2.800\pm0.040$ & $2.425\pm0.014$ & $7.99/11$\\
2(g) & $2.0<y<2.5$ & $1.46\pm0.02$ & $-$ & $2\pm0$ & $1$ & $2.920\pm0.040$ & $2.529\pm0.014$ & $8.79/11$\\
     & $2.5<y<3.0$ & $1.43\pm0.02$ & $-$ & $2\pm0$ & $1$ & $2.860\pm0.040$ & $2.477\pm0.014$ & $9.08/11$\\
     & $3.0<y<3.5$ & $1.40\pm0.02$ & $-$ & $2\pm0$ & $1$ & $2.800\pm0.040$ & $2.425\pm0.014$ & $9.44/11$\\
     & $3.5<y<4.0$ & $1.35\pm0.02$ & $-$ & $2\pm0$ & $1$ & $2.700\pm0.040$ & $2.338\pm0.014$ & $9.93/11$\\
     & $4.0<y<4.5$ & $1.32\pm0.02$ & $-$ & $2\pm0$ & $1$ & $2.640\pm0.040$ & $2.286\pm0.014$ & $14.04/11$\\
2(h) & $2.0<y<2.5$ & $1.80\pm0.02$ & $-$ & $2\pm0$ & $1$ & $3.600\pm0.040$ & $3.118\pm0.014$ & $10.58/11$\\
     & $2.5<y<3.0$ & $1.77\pm0.02$ & $-$ & $2\pm0$ & $1$ & $3.540\pm0.040$ & $3.066\pm0.014$ & $9.08/11$\\
     & $3.0<y<3.5$ & $1.73\pm0.02$ & $-$ & $2\pm0$ & $1$ & $3.460\pm0.040$ & $2.996\pm0.014$ & $9.17/11$\\
     & $3.5<y<4.0$ & $1.60\pm0.02$ & $-$ & $2\pm0$ & $1$ & $3.200\pm0.040$ & $2.771\pm0.014$ & $9.47/11$\\
     & $4.0<y<4.5$ & $1.50\pm0.02$ & $-$ & $2\pm0$ & $1$ & $3.000\pm0.040$ & $2.598\pm0.014$ & $12.09/11$\\
\hline
\end{tabular}}
\end{center}

\newpage

{\scriptsize Table A1. Continued. Values of $\langle
p_t\rangle_1$, $\langle p_t\rangle_2$, $N_1$, $k_E$, $\langle
p_T\rangle$, $T_i$, and $\chi^2$/ndof corresponding to the solid
curves in Figs. 3 and 4. In all cases, $N_2=2$ which is not listed
in the table. In the case of ndof $\leq0$, we use ``$-$" to
mention.} \vspace{-4mm}
\begin{center}
{\scriptsize
\begin{tabular} {ccccccccc}\\ \hline\hline Figure & Main selection & $\langle p_t\rangle_1$ (GeV/$c$) & $\langle p_t\rangle_2$ (GeV/$c$) & $N_1$ & $k_E$ & $\langle p_T\rangle$ (GeV/$c$) & $T_i$ (GeV) & $\chi^2$/ndof\\
\hline
3(a) & 1.8 TeV  & $1.00\pm0.05$ & $2.70\pm0.20$ & $2\pm0$ & $0.98\pm0.01$ & $2.068\pm0.060$ & $1.791\pm0.045$ & $3.80/-$\\
     & 1.96 TeV & $1.30\pm0.05$ & $3.30\pm0.10$ & $2\pm0$ & $0.99\pm0.01$ & $2.640\pm0.064$ & $2.286\pm0.055$ & $11.16/20$\\
3(b) & Inclusive $\psi(2S)$ & $1.58\pm0.10$ & $-$ & $2\pm0$ & $1$ & $3.160\pm0.200$ & $2.737\pm0.071$ & $13.01/2$\\
     & Prompt $\psi(2S)$    & $1.61\pm0.05$ & $-$ & $2\pm0$ & $1$ & $3.220\pm0.100$ & $2.789\pm0.035$ & $5.49/2$\\
     & $\psi(2S)$ from $b$  & $1.65\pm0.10$ & $-$ & $2\pm0$ & $1$ & $3.300\pm0.200$ & $2.858\pm0.071$ & $8.51/2$\\
3(c) & $0.00<|y|<0.75$ & $1.18\pm0.03$ & $3.75\pm0.20$ & $3\pm0$ & $0.98\pm0.01$ & $3.619\pm0.069$ & $2.962\pm0.074$ & $2.39/-$\\
     & $0.75<|y|<1.50$ & $1.14\pm0.02$ & $3.65\pm0.05$ & $3\pm0$ & $0.98\pm0.01$ & $3.498\pm0.056$ & $2.863\pm0.070$ & $3.50/-$\\
     & $1.50<|y|<2.00$ & $1.12\pm0.01$ & $3.60\pm0.05$ & $3\pm0$ & $0.98\pm0.01$ & $3.437\pm0.046$ & $2.813\pm0.069$ & $4.25/-$\\
3(d) & $0.00<|y|<0.75$ & $1.20\pm0.03$ & $3.80\pm0.10$ & $3\pm0$ & $0.97\pm0.01$ & $3.720\pm0.068$ & $3.049\pm0.074$ & $3.23/-$\\
     & $0.75<|y|<1.50$ & $1.16\pm0.03$ & $3.70\pm0.10$ & $3\pm0$ & $0.97\pm0.01$ & $3.598\pm0.067$ & $2.948\pm0.072$ & $7.64/-$\\
     & $1.50<|y|<2.00$ & $1.14\pm0.03$ & $3.65\pm0.10$ & $3\pm0$ & $0.97\pm0.01$ & $3.536\pm0.049$ & $2.898\pm0.071$ & $2.33/-$\\
3(e) & $0<|y|<0.9$     & $3.50\pm0.10$ & $-$ & $1\pm0$ & $1$ & $3.500\pm0.100$ & $3.031\pm0.035$ & $1.45/-$\\
     & $0.9<|y|<1.5$   & $3.30\pm0.10$ & $-$ & $1\pm0$ & $1$ & $3.300\pm0.100$ & $2.858\pm0.035$ & $1.81/-$\\
     & $1.5<|y|<1.93$  & $3.10\pm0.10$ & $-$ & $1\pm0$ & $1$ & $3.100\pm0.100$ & $2.685\pm0.035$ & $2.75/-$\\
     & $1.93<|y|<2.4$  & $3.00\pm0.10$ & $-$ & $1\pm0$ & $1$ & $3.000\pm0.100$ & $2.598\pm0.035$ & $3.50/-$\\
3(f) & $|y|<1.2$       & $2.10\pm0.20$ & $-$ & $2\pm0$ & $1$ & $4.200\pm0.040$ & $3.637\pm0.141$ & $4.51/6$\\
     & $1.2<|y|<1.6$   & $1.90\pm0.20$ & $-$ & $2\pm0$ & $1$ & $3.800\pm0.040$ & $3.291\pm0.141$ & $6.30/4$\\
     & $1.6<|y|<2.4$   & $1.70\pm0.20$ & $-$ & $2\pm0$ & $1$ & $3.400\pm0.040$ & $2.944\pm0.141$ & $5.71/4$\\
3(g) & $|y|<1.2$       & $2.60\pm0.20$ & $-$ & $2\pm0$ & $1$ & $5.200\pm0.040$ & $4.503\pm0.141$ & $4.19/6$\\
     & $1.2<|y|<1.6$   & $2.10\pm0.20$ & $-$ & $2\pm0$ & $1$ & $4.200\pm0.040$ & $3.637\pm0.141$ & $6.61/4$\\
     & $1.6<|y|<2.4$   & $1.90\pm0.20$ & $-$ & $2\pm0$ & $1$ & $3.800\pm0.040$ & $3.291\pm0.141$ & $5.58/4$\\
\hline
4(a) & $|y|<0.75$     & $4.90\pm0.20$ & $-$ & $1\pm0$  & $1$  & $4.900\pm0.200$ & $4.900\pm0.141$ & $16.68/7$\\
     & $0.75<|y|<1.5$ & $4.85\pm0.20$ & $-$ & $1\pm0$  & $1$  & $4.850\pm0.200$ & $4.850\pm0.141$ & $12.64/7$\\
     & $1.5<|y|<2.0$  & $4.60\pm0.20$ & $-$ & $1\pm0$  & $1$  & $4.600\pm0.200$ & $4.600\pm0.141$ & $14.92/7$\\
4(b) & $|y|<0.75$     & $5.70\pm0.20$ & $-$ & $1\pm0$  & $1$  & $5.700\pm0.200$ & $5.700\pm0.141$ & $9.50/7$\\
     & $0.75<|y|<1.5$ & $5.60\pm0.20$ & $-$ & $1\pm0$  & $1$  & $5.600\pm0.200$ & $5.600\pm0.141$ & $12.60/7$\\
     & $1.5<|y|<2.0$  & $4.90\pm0.20$ & $-$ & $1\pm0$  & $1$  & $4.900\pm0.200$ & $4.900\pm0.141$ & $23.39/7$\\
4(c) & Prompt $\psi(2S)$     & $4.50\pm0.50$ & $-$ & $1\pm0$  & $1$  & $4.500\pm0.500$ & $4.500\pm0.354$ & $31.00/2$\\
     & Non-prompt $\psi(2S)$ & $5.00\pm0.50$ & $-$ & $1\pm0$  & $1$  & $5.000\pm0.500$ & $5.000\pm0.354$ & $24.00/2$\\
4(d) & $0.0<|y|<0.3$  & $6.00\pm0.20$ & $-$ & $1\pm0$¡¡& $1$  & $6.000\pm0.200$ & $6.000\pm0.141$ & $78.14/6$\\
     & $0.3<|y|<0.6$  & $5.90\pm0.20$ & $-$ & $1\pm0$  & $1$  & $5.900\pm0.200$ & $5.900\pm0.141$ & $149.64/6$\\
     & $0.6<|y|<0.9$  & $5.80\pm0.20$ & $-$ & $1\pm0$  & $1$  & $5.800\pm0.200$ & $5.800\pm0.141$ & $191.75/6$\\
     & $0.9<|y|<1.2$  & $5.70\pm0.20$ & $-$ & $1\pm0$  & $1$  & $5.700\pm0.200$ & $5.700\pm0.141$ & $444.14/6$\\
\hline
\end{tabular}}
\end{center}

\newpage

{\scriptsize Table A1. Continued. Values of $\langle
p_t\rangle_1$, $\langle p_t\rangle_2$, $N_1$, $k_E$, $\langle
p_T\rangle$, $T_i$, and $\chi^2$/ndof corresponding to the solid
curves in Figs. 5 and 6. In all cases, $N_2=2$ which is not listed
in the table. In the case of ndof $\leq0$, we use ``$-$" to
mention.} \vspace{-4mm}
\begin{center}
{\scriptsize
\begin{tabular} {ccccccccc}\\ \hline\hline Figure & Main selection & $\langle p_t\rangle_1$ (GeV/$c$) & $\langle p_t\rangle_2$ (GeV/$c$) & $N_1$ & $k_E$ & $\langle p_T\rangle$ (GeV/$c$) & $T_i$ (GeV) & $\chi^2$/ndof\\
\hline
5(a) & $\Upsilon(1S)$ & $2.35\pm0.10$ & $-$ & $2\pm0$ & $1$ & $4.700\pm0.200$ & $4.070\pm0.071$ & $14.28/11$\\
     & $\Upsilon(2S)$ & $2.65\pm0.10$ & $-$ & $2\pm0$ & $1$ & $5.300\pm0.200$ & $4.590\pm0.071$ & $11.50/6$\\
     & $\Upsilon(3S)$ & $2.70\pm0.10$ & $-$ & $2\pm0$ & $1$ & $5.400\pm0.200$ & $4.677\pm0.071$ & $7.99/6$\\
5(b) & $\Upsilon(1S)$ & $2.35\pm0.10$ & $-$ & $2\pm0$ & $1$ & $4.700\pm0.200$ & $4.070\pm0.071$ & $7.79/3$\\
     & $\Upsilon(2S)$ & $2.65\pm0.10$ & $-$ & $2\pm0$ & $1$ & $5.300\pm0.200$ & $4.590\pm0.071$ & $4.25/3$\\
     & $\Upsilon(3S)$ & $2.75\pm0.10$ & $-$ & $2\pm0$ & $1$ & $5.500\pm0.200$ & $4.763\pm0.071$ & $2.36/3$\\
5(c) & $\Upsilon(1S)$ & $2.75\pm0.10$ & $-$ & $2\pm0$ & $1$ & $5.500\pm0.200$ & $4.763\pm0.071$ & $13.89/3$\\
     & $\Upsilon(2S)$ & $3.04\pm0.10$ & $-$ & $2\pm0$ & $1$ & $6.080\pm0.200$ & $5.265\pm0.071$ & $11.00/-$\\
     & $\Upsilon(3S)$ & $3.28\pm0.10$ & $-$ & $2\pm0$ & $1$ & $6.540\pm0.200$ & $5.681\pm0.071$ & $12.00/-$\\
5(d) & $2.0<y<2.5$ & $2.87\pm0.10$ & $-$ & $2\pm0$ & $1$ & $5.740\pm0.200$ & $4.971\pm0.071$ & $10.30/21$\\
     & $2.5<y<3.0$ & $2.74\pm0.10$ & $-$ & $2\pm0$ & $1$ & $5.480\pm0.200$ & $4.746\pm0.071$ & $12.57/21$\\
     & $3.0<y<3.5$ & $2.66\pm0.10$ & $-$ & $2\pm0$ & $1$ & $5.320\pm0.200$ & $4.607\pm0.071$ & $13.98/21$\\
     & $3.5<y<4.0$ & $2.30\pm0.10$ & $-$ & $2\pm0$ & $1$ & $4.600\pm0.200$ & $3.984\pm0.071$ & $11.90/18$\\
     & $4.0<y<4.5$ & $2.15\pm0.10$ & $-$ & $2\pm0$ & $1$ & $4.300\pm0.200$ & $3.724\pm0.071$ & $8.28/12$\\
5(e) & $2.0<y<2.5$ & $3.24\pm0.10$ & $-$ & $2\pm0$ & $1$ & $6.480\pm0.200$ & $5.612\pm0.071$ & $14.04/21$\\
     & $2.5<y<3.0$ & $3.07\pm0.10$ & $-$ & $2\pm0$ & $1$ & $6.140\pm0.200$ & $5.317\pm0.071$ & $11.76/21$\\
     & $3.0<y<3.5$ & $2.98\pm0.10$ & $-$ & $2\pm0$ & $1$ & $5.960\pm0.200$ & $5.162\pm0.071$ & $10.98/21$\\
     & $3.5<y<4.0$ & $2.60\pm0.10$ & $-$ & $2\pm0$ & $1$ & $5.200\pm0.200$ & $4.503\pm0.071$ & $9.67/18$\\
     & $4.0<y<4.5$ & $2.30\pm0.10$ & $-$ & $2\pm0$ & $1$ & $4.600\pm0.200$ & $3.984\pm0.071$ & $9.78/12$\\
5(f) & $2.0<y<2.5$ & $3.58\pm0.10$ & $-$ & $2\pm0$ & $1$ & $7.160\pm0.200$ & $6.201\pm0.071$ & $10.56/21$\\
     & $2.5<y<3.0$ & $3.36\pm0.10$ & $-$ & $2\pm0$ & $1$ & $6.720\pm0.200$ & $5.820\pm0.071$ & $8.31/21$\\
     & $3.0<y<3.5$ & $3.15\pm0.10$ & $-$ & $2\pm0$ & $1$ & $6.300\pm0.200$ & $5.456\pm0.071$ & $10.95/21$\\
     & $3.5<y<4.0$ & $2.80\pm0.10$ & $-$ & $2\pm0$ & $1$ & $5.600\pm0.200$ & $4.850\pm0.071$ & $10.59/18$\\
     & $4.0<y<4.5$ & $2.50\pm0.10$ & $-$ & $2\pm0$ & $1$ & $5.000\pm0.200$ & $4.330\pm0.071$ & $9.28/12$\\
\hline
6(a) & $2.0<y<2.5$ & $2.93\pm0.10$ & $-$ & $2\pm0$ & $1$ & $5.860\pm0.200$ & $5.075\pm0.071$ & $13.98/21$\\
     & $2.5<y<3.0$ & $2.80\pm0.10$ & $-$ & $2\pm0$ & $1$ & $5.600\pm0.200$ & $4.850\pm0.071$ & $17.21/21$\\
     & $3.0<y<3.5$ & $2.71\pm0.10$ & $-$ & $2\pm0$ & $1$ & $5.420\pm0.200$ & $4.694\pm0.071$ & $15.08/21$\\
     & $3.5<y<4.0$ & $2.50\pm0.10$ & $-$ & $2\pm0$ & $1$ & $5.000\pm0.200$ & $4.330\pm0.071$ & $7.56/18$\\
     & $4.0<y<4.5$ & $2.18\pm0.10$ & $-$ & $2\pm0$ & $1$ & $4.360\pm0.200$ & $3.776\pm0.071$ & $8.70/12$\\
6(b) & $2.0<y<2.5$ & $3.25\pm0.10$ & $-$ & $2\pm0$ & $1$ & $6.500\pm0.200$ & $5.629\pm0.071$ & $13.31/21$\\
     & $2.5<y<3.0$ & $3.18\pm0.10$ & $-$ & $2\pm0$ & $1$ & $6.360\pm0.200$ & $5.508\pm0.071$ & $15.28/21$\\
     & $3.0<y<3.5$ & $3.00\pm0.10$ & $-$ & $2\pm0$ & $1$ & $6.000\pm0.200$ & $5.196\pm0.071$ & $12.05/21$\\
     & $3.5<y<4.0$ & $2.80\pm0.10$ & $-$ & $2\pm0$ & $1$ & $5.600\pm0.200$ & $4.850\pm0.071$ & $10.23/18$\\
     & $4.0<y<4.5$ & $2.40\pm0.10$ & $-$ & $2\pm0$ & $1$ & $4.800\pm0.200$ & $4.157\pm0.071$ & $9.45/12$\\
6(c) & $2.0<y<2.5$ & $3.65\pm0.10$ & $-$ & $2\pm0$ & $1$ & $7.300\pm0.200$ & $6.322\pm0.071$ & $10.23/21$\\
     & $2.5<y<3.0$ & $3.47\pm0.10$ & $-$ & $2\pm0$ & $1$ & $6.940\pm0.200$ & $6.010\pm0.071$ & $12.87/21$\\
     & $3.0<y<3.5$ & $3.30\pm0.10$ & $-$ & $2\pm0$ & $1$ & $6.600\pm0.200$ & $5.716\pm0.071$ & $9.25/21$\\
     & $3.5<y<4.0$ & $3.00\pm0.10$ & $-$ & $2\pm0$ & $1$ & $6.000\pm0.200$ & $5.196\pm0.071$ & $7.25/18$\\
     & $4.0<y<4.5$ & $2.60\pm0.10$ & $-$ & $2\pm0$ & $1$ & $5.200\pm0.200$ & $4.503\pm0.071$ & $9.06/12$\\
6(d) & $1.5<y<2.0$ & $3.00\pm0.10$ & $-$ & $2\pm0$ & $1$ & $6.000\pm0.200$ & $5.196\pm0.071$ & $4.47/4$\\
     & $2.0<y<2.5$ & $2.95\pm0.10$ & $-$ & $2\pm0$ & $1$ & $5.900\pm0.200$ & $5.110\pm0.083$ & $2.43/4$\\
     & $2.5<y<3.0$ & $2.85\pm0.10$ & $-$ & $2\pm0$ & $1$ & $5.700\pm0.200$ & $4.936\pm0.082$ & $3.25/4$\\
     & $3.0<y<3.5$ & $2.75\pm0.10$ & $-$ & $2\pm0$ & $1$ & $5.500\pm0.200$ & $4.763\pm0.081$ & $3.64/4$\\
     & $3.5<y<4.0$ & $2.60\pm0.10$ & $-$ & $2\pm0$ & $1$ & $5.200\pm0.200$ & $4.503\pm0.071$ & $4.52/3$\\
6(e) & $2.0<y<2.5$ & $3.25\pm0.10$ & $-$ & $2\pm0$ & $1$ & $6.500\pm0.200$ & $5.629\pm0.071$ & $18.79/21$\\
     & $2.5<y<3.0$ & $3.00\pm0.10$ & $-$ & $2\pm0$ & $1$ & $6.000\pm0.200$ & $5.196\pm0.071$ & $17.87/22$\\
     & $3.0<y<3.5$ & $2.80\pm0.10$ & $-$ & $2\pm0$ & $1$ & $5.600\pm0.200$ & $4.850\pm0.071$ & $18.85/21$\\
     & $3.5<y<4.0$ & $2.75\pm0.10$ & $-$ & $2\pm0$ & $1$ & $5.500\pm0.200$ & $4.763\pm0.071$ & $16.42/19$\\
     & $4.0<y<4.5$ & $2.70\pm0.10$ & $-$ & $2\pm0$ & $1$ & $5.400\pm0.200$ & $4.677\pm0.071$ & $11.03/13$\\
6(f) & $2.0<y<2.5$ & $3.35\pm0.10$ & $-$ & $2\pm0$ & $1$ & $6.700\pm0.200$ & $5.802\pm0.071$ & $17.71/21$\\
     & $2.5<y<3.0$ & $3.30\pm0.10$ & $-$ & $2\pm0$ & $1$ & $6.600\pm0.200$ & $5.716\pm0.071$ & $23.75/22$\\
     & $3.0<y<3.5$ & $3.20\pm0.10$ & $-$ & $2\pm0$ & $1$ & $6.400\pm0.200$ & $5.543\pm0.071$ & $16.94/21$\\
     & $3.5<y<4.0$ & $3.10\pm0.10$ & $-$ & $2\pm0$ & $1$ & $6.200\pm0.200$ & $5.369\pm0.071$ & $20.21/19$\\
     & $4.0<y<4.5$ & $3.00\pm0.10$ & $-$ & $2\pm0$ & $1$ & $6.000\pm0.200$ & $5.196\pm0.071$ & $10.58/13$\\
6(g) & $2.0<y<2.5$ & $3.80\pm0.10$ & $-$ & $2\pm0$ & $1$ & $7.600\pm0.200$ & $6.582\pm0.071$ & $13.87/21$\\
     & $2.5<y<3.0$ & $3.60\pm0.10$ & $-$ & $2\pm0$ & $1$ & $7.200\pm0.200$ & $6.235\pm0.071$ & $14.69/22$\\
     & $3.0<y<3.5$ & $3.50\pm0.10$ & $-$ & $2\pm0$ & $1$ & $7.000\pm0.200$ & $6.062\pm0.071$ & $12.10/21$\\
     & $3.5<y<4.0$ & $3.30\pm0.10$ & $-$ & $2\pm0$ & $1$ & $6.600\pm0.200$ & $5.716\pm0.071$ & $16.61/19$\\
     & $4.0<y<4.5$ & $3.20\pm0.10$ & $-$ & $2\pm0$ & $1$ & $6.400\pm0.200$ & $5.543\pm0.071$ & $11.17/13$\\
\hline
\end{tabular}}
\end{center}

\newpage

{\scriptsize Table A2. Values of $p_0$, $n_0$, and $\chi^2$/ndof
corresponding to the dashed curves in Figs. 1 and 2, as well as
values of $T$, $n$, and $\chi^2$/ndof corresponding to the dotted
curves in Figs. 1 and 2.} \vspace{-4mm}
\begin{center}
{\scriptsize
\begin{tabular} {ccccccccc}\\ \hline\hline Figure & Main selection & $p_0$ (GeV/$c$) & $n_0$ & $\chi^2$/ndof & $T$ (GeV) & $n$ & $\chi^2$/ndof\\
\hline
1(a) & 0--20\%  & $6.20\pm0.50$ & $7.80\pm0.50$ & $1.07/-$    & $0.39\pm0.02$ & $5.70\pm1.00$  & $0.80/-$\\
     & 20--40\% & $6.10\pm0.50$ & $7.90\pm0.50$ & $2.11/-$    & $0.37\pm0.02$ & $6.00\pm1.00$  & $1.01/-$\\
     & 40--60\% & $5.95\pm0.50$ & $8.05\pm0.50$ & $0.23/-$    & $0.36\pm0.02$ & $6.20\pm1.00$  & $1.11/-$\\
     & 0--60\%  & $6.00\pm0.50$ & $8.00\pm0.50$ & $0.59/-$    & $0.37\pm0.02$ & $6.00\pm1.00$  & $0.44/-$\\
1(b) & 0--20\%  & $6.60\pm0.50$ & $8.00\pm0.50$ & $3.00/-$    & $0.41\pm0.02$ & $5.90\pm1.00$  & $1.40/-$\\
     & 20--40\% & $6.50\pm0.50$ & $8.10\pm0.50$ & $1.73/-$    & $0.39\pm0.02$ & $6.20\pm1.00$  & $1.07/-$\\
     & 40--60\% & $6.35\pm0.50$ & $8.25\pm0.50$ & $2.16/-$    & $0.38\pm0.02$ & $6.30\pm1.00$  & $0.78/-$\\
     & 0--60\%  & $6.40\pm0.30$ & $8.20\pm0.30$ & $2.94/-$    & $0.39\pm0.02$ & $6.20\pm1.00$  & $1.37/-$\\
1(c) & 0--20\%  & $7.00\pm0.50$ & $9.30\pm0.50$ & $2.36/2$    & $0.43\pm0.02$ & $6.10\pm1.00$  & $0.84/2$\\
     & 20--40\% & $6.80\pm0.20$ & $9.50\pm0.20$ & $4.43/2$    & $0.40\pm0.02$ & $6.30\pm1.00$  & $1.86/2$\\
     & 40--60\% & $6.70\pm0.50$ & $9.60\pm0.50$ & $3.85/2$    & $0.39\pm0.02$ & $6.40\pm1.00$  & $1.46/2$\\
     & 0--60\%  & $6.80\pm0.50$ & $9.50\pm0.50$ & $3.86/2$    & $0.40\pm0.02$ & $6.30\pm1.00$  & $0.41/2$\\
1(d) & Full cross section     & $6.80\pm0.30$ & $11.50\pm0.30$& $35.57/16$ & $0.44\pm0.02$ & $8.00\pm0.20$ & $24.30/16$\\
     & Fiducial cross section & $7.00\pm0.30$ & $11.00\pm0.30$& $70.35/16$ & $0.47\pm0.01$ & $7.50\pm0.20$ & $54.75/16$\\
1(e) & Full cross section     & $7.50\pm0.50$ & $11.70\pm1.00$& $11.89/2$  & $0.45\pm0.02$ & $8.10\pm1.00$ & $2.22/2$\\
     & Fiducial cross section & $8.00\pm0.30$ & $11.20\pm0.50$& $4.31/2$   & $0.50\pm0.03$ & $7.60\pm1.00$ & $10.25/2$\\
1(f) & 1.8 TeV  & $8.10\pm0.30$  & $11.80\pm0.30$ & $4.87/8$    & $0.55\pm0.02$ & $8.50\pm0.20$  & $4.20/8$\\
     & 1.96 TeV & $8.20\pm0.30$  & $12.00\pm0.50$ & $26.09/21$  & $0.57\pm0.03$ & $8.90\pm0.50$  & $21.40/21$\\
1(g) & 0--20\%  & $12.40\pm0.50$ & $11.50\pm0.50$ & $217.00/10$ & $0.53\pm0.05$ & $11.50\pm2.00$ & $11.31/10$\\
     & 20--40\% & $12.20\pm0.50$ & $11.80\pm0.50$ & $90.75/10$  & $0.52\pm0.05$ & $11.60\pm2.00$ & $10.92/10$\\
     & 40--90\% & $12.00\pm0.50$ & $12.00\pm0.50$ & $41.80/10$  & $0.52\pm0.05$ & $11.65\pm2.00$ & $15.71/10$\\
\hline
2(a) & $1.5<y<2.0$ & $13.00\pm0.50$ & $12.50\pm0.50$ & $8.97/5$   & $0.85\pm0.05$ & $10.00\pm2.00$ & $2.16/5$\\
     & $2.0<y<2.5$ & $12.90\pm0.50$ & $12.60\pm0.50$ & $8.52/5$   & $0.84\pm0.05$ & $10.50\pm2.00$ & $2.66/5$\\
     & $2.5<y<3.0$ & $12.60\pm0.50$ & $12.90\pm0.50$ & $5.91/5$   & $0.83\pm0.05$ & $11.00\pm2.00$ & $2.46/5$\\
     & $3.0<y<3.5$ & $12.40\pm0.50$ & $13.10\pm0.50$ & $5.91/5$   & $0.82\pm0.05$ & $11.50\pm2.00$ & $3.10/5$\\
     & $3.5<y<4.0$ & $12.10\pm0.50$ & $13.50\pm0.50$ & $5.85/5$   & $0.80\pm0.05$ & $12.50\pm2.00$ & $3.61/5$\\
2(b) & $1.5<y<2.0$ & $14.50\pm0.50$ & $12.00\pm0.50$ & $3.10/5$   & $1.00\pm0.10$ & $9.50\pm2.00$  & $3.65/5$\\
     & $2.0<y<2.5$ & $14.40\pm0.50$ & $12.10\pm0.50$ & $3.75/5$   & $0.98\pm0.05$ & $9.70\pm2.00$  & $3.62/5$\\
     & $2.5<y<3.0$ & $14.30\pm0.50$ & $12.20\pm0.50$ & $6.39/5$   & $0.95\pm0.05$ & $10.00\pm2.00$ & $4.60/5$\\
     & $3.0<y<3.5$ & $14.10\pm0.50$ & $12.40\pm0.50$ & $6.75/5$   & $0.93\pm0.05$ & $10.20\pm2.00$ & $6.60/5$\\
     & $3.5<y<4.0$ & $13.50\pm0.50$ & $13.00\pm0.50$ & $11.66/5$  & $0.87\pm0.05$ & $11.00\pm2.00$ & $5.21/5$\\
2(c) & $2.0<y<2.5$ & $13.50\pm0.50$ & $14.00\pm0.50$ & $8.91/11$  & $0.85\pm0.05$ & $12.50\pm2.00$ & $7.93/11$\\
     & $2.5<y<3.0$ & $13.30\pm0.50$ & $14.20\pm0.50$ & $8.66/11$  & $0.84\pm0.05$ & $13.00\pm2.00$ & $9.52/11$\\
     & $3.0<y<3.5$ & $13.00\pm0.50$ & $14.40\pm0.50$ & $8.41/11$  & $0.82\pm0.05$ & $14.00\pm2.00$ & $9.55/11$\\
     & $3.5<y<4.0$ & $12.60\pm0.50$ & $14.80\pm0.50$ & $8.30/10$  & $0.80\pm0.05$ & $15.00\pm2.00$ & $8.99/10$\\
     & $4.0<y<4.5$ & $12.20\pm0.50$ & $15.00\pm0.50$ & $7.80/8$   & $0.74\pm0.05$ & $16.00\pm2.00$ & $4.91/8$\\
2(d) & $2.0<y<2.5$ & $15.00\pm0.50$ & $13.00\pm0.50$ & $3.19/11$  & $1.02\pm0.05$ & $11.00\pm2.00$ & $5.85/11$\\
     & $2.5<y<3.0$ & $14.80\pm0.50$ & $13.20\pm0.50$ & $4.93/11$  & $1.00\pm0.05$ & $11.50\pm2.00$ & $6.44/11$\\
     & $3.0<y<3.5$ & $14.30\pm0.50$ & $13.50\pm0.50$ & $5.32/11$  & $0.98\pm0.05$ & $12.00\pm2.00$ & $5.98/11$\\
     & $3.5<y<4.0$ & $14.10\pm0.50$ & $14.00\pm0.50$ & $8.86/10$  & $0.93\pm0.05$ & $13.00\pm2.00$ & $6.55/10$\\
     & $4.0<y<4.5$ & $14.00\pm0.50$ & $14.20\pm0.50$ & $9.91/8$   & $0.90\pm0.05$ & $13.50\pm2.00$ & $5.26/8$\\
2(e) & $2.0<y<2.5$ & $13.70\pm0.50$ & $14.20\pm0.50$ & $10.12/11$ & $0.88\pm0.05$ & $13.00\pm2.00$ & $7.17/11$\\
     & $2.5<y<3.0$ & $13.50\pm0.50$ & $14.40\pm0.50$ & $5.75/11$  & $0.87\pm0.05$ & $13.50\pm2.00$ & $8.75/11$\\
     & $3.0<y<3.5$ & $13.30\pm0.50$ & $14.60\pm0.50$ & $5.78/11$  & $0.86\pm0.05$ & $14.00\pm2.00$ & $8.16/11$\\
     & $3.5<y<4.0$ & $13.10\pm0.50$ & $14.80\pm0.50$ & $8.76/11$  & $0.84\pm0.05$ & $15.00\pm2.00$ & $10.56/11$\\
     & $4.0<y<4.5$ & $12.50\pm0.50$ & $15.20\pm0.50$ & $9.92/11$  & $0.77\pm0.05$ & $16.50\pm2.00$ & $7.25/11$\\
2(f) & $2.0<y<2.5$ & $15.60\pm0.50$ & $13.10\pm0.50$ & $5.45/11$  & $1.07\pm0.05$ & $11.50\pm2.00$ & $5.17/11$\\
     & $2.5<y<3.0$ & $15.40\pm0.50$ & $13.30\pm0.50$ & $8.64/11$  & $1.05\pm0.05$ & $12.00\pm2.00$ & $6.79/11$\\
     & $3.0<y<3.5$ & $15.00\pm0.50$ & $13.70\pm0.50$ & $5.17/11$  & $1.02\pm0.05$ & $12.50\pm2.00$ & $7.16/11$\\
     & $3.5<y<4.0$ & $14.70\pm0.50$ & $14.10\pm0.50$ & $11.56/11$ & $0.98\pm0.05$ & $13.50\pm2.00$ & $7.07/11$\\
     & $4.0<y<4.5$ & $14.20\pm0.50$ & $14.40\pm0.50$ & $9.57/11$  & $0.93\pm0.05$ & $14.00\pm2.00$ & $5.81/11$\\
2(g) & $2.0<y<2.5$ & $15.30\pm0.50$ & $14.80\pm0.50$ & $9.38/11$  & $0.93\pm0.05$ & $14.00\pm2.00$ & $6.16/11$\\
     & $2.5<y<3.0$ & $15.20\pm0.50$ & $14.90\pm0.50$ & $8.95/11$  & $0.91\pm0.05$ & $14.50\pm2.00$ & $8.12/11$\\
     & $3.0<y<3.5$ & $15.00\pm0.50$ & $15.00\pm0.50$ & $7.32/11$  & $0.89\pm0.05$ & $14.80\pm2.00$ & $9.17/11$\\
     & $3.5<y<4.0$ & $14.40\pm0.50$ & $15.30\pm0.50$ & $11.22/11$ & $0.86\pm0.05$ & $15.30\pm2.00$ & $9.72/11$\\
     & $4.0<y<4.5$ & $14.20\pm0.50$ & $15.40\pm0.50$ & $8.86/11$  & $0.84\pm0.05$ & $16.80\pm2.00$ & $13.48/11$\\
2(h) & $2.0<y<2.5$ & $18.00\pm0.50$ & $13.80\pm0.50$ & $10.77/11$ & $1.20\pm0.10$ & $12.00\pm2.00$ & $8.03/11$\\
     & $2.5<y<3.0$ & $17.80\pm0.50$ & $13.90\pm0.50$ & $5.92/11$  & $1.15\pm0.10$ & $12.50\pm2.00$ & $8.62/11$\\
     & $3.0<y<3.5$ & $17.60\pm0.50$ & $14.00\pm0.50$ & $8.20/11$  & $1.12\pm0.10$ & $12.80\pm2.00$ & $10.17/11$\\
     & $3.5<y<4.0$ & $16.50\pm0.50$ & $14.40\pm0.50$ & $9.56/11$  & $1.05\pm0.10$ & $14.20\pm2.00$ & $8.75/11$\\
     & $4.0<y<4.5$ & $15.50\pm0.50$ & $14.60\pm0.50$ & $13.83/11$ & $1.00\pm0.05$ & $14.70\pm2.00$ & $10.58/11$\\
\hline
\end{tabular}}
\end{center}

\newpage

{\scriptsize Table A2. Continued. Values of $p_0$, $n_0$, and
$\chi^2$/ndof corresponding to the dashed curves in Figs. 3 and 4,
as well as values of $T$, $n$, and $\chi^2$/ndof corresponding to
the dotted curves in Figs. 3 and 4. The parameter values in the
first (second) row in each panel for Fig. 4(d) is for the first
(second) component, where * denotes $k_H$ and ** denotes $k_L$. In
other cases only the single component is used.} \vspace{-4mm}
\begin{center}
{\scriptsize
\begin{tabular} {ccccccccc}\\ \hline\hline Figure & Main selection & $p_0$ (GeV/$c$) & $n_0$ & $\chi^2$/ndof & $T$ (GeV) & $n$ & $\chi^2$/ndof\\
\hline
3(a) & 1.8 TeV     & $9.00\pm0.50$  & $12.10\pm0.50$ & $4.00/2$   & $0.65\pm0.05$ & $9.00\pm1.00$  & $5.50/2$\\
     & 1.96 TeV    & $10.30\pm0.50$ & $12.20\pm0.50$ & $17.50/22$ & $0.75\pm0.05$ & $9.30\pm1.00$  & $9.99/22$\\
3(b) & Inclusive $\psi(2S)$ & $14.10\pm0.50$ & $13.10\pm0.50$ & $19.96/2$  & $1.07\pm0.10$ & $12.30\pm2.00$ & $4.51/2$\\
     & Prompt $\psi(2S)$    & $14.20\pm0.50$ & $13.00\pm0.50$ & $10.43/2$  & $1.10\pm0.10$ & $12.00\pm1.00$ & $2.36/2$\\
     & $\psi(2S)$ from $b$  & $14.30\pm0.50$ & $12.80\pm0.50$ & $15.75/2$  & $1.13\pm0.10$ & $11.70\pm2.00$ & $7.59/2$\\
3(c) & $0.00<|y|<0.75$ & $14.50\pm0.50$ & $13.50\pm0.50$ & $7.75/2$  & $1.16\pm0.10$ & $12.80\pm1.00$ & $7.00/2$\\
     & $0.75<|y|<1.50$ & $14.30\pm0.50$ & $13.60\pm0.50$ & $5.14/2$  & $1.14\pm0.10$ & $13.00\pm1.00$ & $3.22/2$\\
     & $1.50<|y|<2.00$ & $14.10\pm0.50$ & $13.70\pm0.50$ & $3.14/2$  & $1.12\pm0.10$ & $13.20\pm1.00$ & $3.14/2$\\
3(d) & $0.00<|y|<0.75$ & $16.00\pm0.50$ & $13.00\pm0.50$ & $6.25/2$  & $1.32\pm0.10$ & $12.50\pm1.00$ & $4.33/2$\\
     & $0.75<|y|<1.50$ & $15.90\pm0.50$ & $13.10\pm0.50$ & $9.89/2$  & $1.31\pm0.10$ & $12.60\pm1.00$ & $7.39/2$\\
     & $1.50<|y|<2.00$ & $15.80\pm0.50$ & $13.20\pm0.50$ & $11.25/2$ & $1.30\pm0.10$ & $12.70\pm1.00$ & $10.25/2$\\
3(e) & $0<|y|<0.9$     & $15.50\pm0.50$ & $11.20\pm0.50$ & $2.00/-$  & $1.50\pm0.05$ & $12.00\pm1.00$ & $2.21/-$\\
     & $0.9<|y|<1.5$   & $15.30\pm0.50$ & $11.30\pm0.50$ & $1.85/-$  & $1.48\pm0.05$ & $12.20\pm1.00$ & $2.69/-$\\
     & $1.5<|y|<1.93$  & $15.10\pm0.50$ & $11.40\pm0.50$ & $2.75/-$  & $1.45\pm0.05$ & $12.50\pm1.00$ & $3.50/-$\\
     & $1.93<|y|<2.4$  & $14.90\pm0.50$ & $11.50\pm0.50$ & $3.50/-$  & $1.42\pm0.05$ & $12.70\pm1.00$ & $4.84/-$\\
3(f) & $|y|<1.2$       & $16.00\pm0.50$ & $14.50\pm0.50$ & $2.39/6$  & $1.50\pm0.10$ & $14.50\pm1.00$ & $3.39/6$\\
     & $1.2<|y|<1.6$   & $15.50\pm0.50$ & $14.80\pm0.50$ & $3.80/4$  & $1.38\pm0.10$ & $15.00\pm1.00$ & $6.67/4$\\
     & $1.6<|y|<2.4$   & $15.00\pm0.50$ & $15.10\pm0.50$ & $2.10/4$  & $1.28\pm0.10$ & $15.50\pm1.00$ & $3.64/4$\\
3(g) & $|y|<1.2$       & $18.00\pm0.50$ & $13.00\pm0.50$ & $4.17/6$  & $1.65\pm0.05$ & $14.00\pm1.00$ & $3.98/6$\\
     & $1.2<|y|<1.6$   & $17.00\pm0.50$ & $13.50\pm0.50$ & $4.23/4$  & $1.57\pm0.05$ & $14.50\pm1.00$ & $4.86/4$\\
     & $1.6<|y|<2.4$   & $16.00\pm0.50$ & $14.00\pm0.50$ & $3.49/4$  & $1.50\pm0.05$ & $15.00\pm1.00$ & $2.89/4$\\
\hline
4(a) & $|y|<0.75$      & $25.00\pm0.50$ & $14.00\pm0.50$ & $6.84/7$  & $1.85\pm0.05$ & $14.00\pm1.00$ & $7.88/7$\\
     & $0.75<|y|<1.5$  & $24.60\pm0.50$ & $14.20\pm0.50$ & $6.81/7$  & $1.82\pm0.05$ & $14.30\pm1.00$ & $7.69/7$\\
     & $1.5<|y|<2.0$   & $24.20\pm0.50$ & $14.40\pm0.50$ & $6.64/7$  & $1.79\pm0.10$ & $14.60\pm1.00$ & $6.67/7$\\
4(b) & $|y|<0.75$      & $25.50\pm0.50$ & $12.50\pm0.50$ & $7.78/7$  & $2.20\pm0.10$ & $13.00\pm1.00$ & $8.77/7$\\
     & $0.75<|y|<1.5$  & $25.30\pm0.50$ & $12.60\pm0.50$ & $5.70/7$  & $2.10\pm0.10$ & $13.20\pm1.00$ & $8.45/7$\\
     & $1.5<|y|<2.0$   & $24.50\pm0.50$ & $13.00\pm0.50$ & $8.56/7$  & $2.00\pm0.10$ & $13.40\pm1.00$ & $7.03/7$\\
4(c) & Prompt $\psi(2S)$     & $26.00\pm0.50$ & $14.60\pm0.50$ & $16.25/2$& $1.90\pm0.05$ & $14.50\pm1.00$ & $17.27/2$\\
     & Non-prompt $\psi(2S)$ & $26.50\pm0.50$ & $13.00\pm0.50$ & $9.40/2$ & $2.30\pm0.10$ & $13.50\pm1.00$ & $8.59/2$\\
4(d) & $0.0<|y|<0.3$  & $35.00\pm1.00$ & $18.00\pm1.00$ & $0.80\pm0.05$* & $2.35\pm0.20$ & $18.00\pm2.00$ & $0.75\pm0.05$**\\
     &                & $20.00\pm1.00$ & $9.00\pm0.50$  & $3.94/3$       & $2.35\pm0.10$ & $8.40\pm0.50$  & $4.17/3$\\
     & $0.3<|y|<0.6$  & $34.50\pm1.00$ & $18.50\pm1.00$ & $0.80\pm0.05$* & $2.30\pm0.20$ & $18.50\pm2.00$ & $0.75\pm0.05$**\\
     &                & $20.00\pm1.00$ & $9.00\pm0.20$  & $4.08/3$       & $2.30\pm0.10$ & $8.20\pm0.50$  & $4.47/3$\\
     & $0.6<|y|<0.9$  & $34.00\pm1.00$ & $19.00\pm1.00$ & $0.80\pm0.05$* & $2.25\pm0.20$ & $19.00\pm2.00$ & $0.75\pm0.05$**\\
     &                & $20.00\pm1.00$ & $9.00\pm0.20$  & $7.50/3$       & $2.25\pm0.10$ & $8.00\pm0.50$  & $6.95/3$\\
     & $0.9<|y|<1.2$  & $33.50\pm1.00$ & $19.50\pm1.00$ & $0.80\pm0.05$* & $2.20\pm0.20$ & $19.50\pm2.00$ & $0.75\pm0.05$**\\
     &                & $20.00\pm1.00$ & $9.00\pm0.20$  & $9.81/3$       & $2.20\pm0.10$ & $7.80\pm0.30$  & $6.12/3$\\
\hline
\end{tabular}}
\end{center}

\newpage

{\scriptsize Table A2. Continued. Values of $p_0$, $n_0$, and
$\chi^2$/ndof corresponding to the dashed curves in Figs. 5 and 6,
as well as values of $T$, $n$, and $\chi^2$/ndof corresponding to
the dotted curves in Figs. 5 and 6.} \vspace{-4mm}
\begin{center}
{\scriptsize
\begin{tabular} {ccccccccc}\\ \hline\hline Figure & Main selection & $p_0$ (GeV/$c$) & $n_0$ & $\chi^2$/ndof & $T$ (GeV) & $n$ & $\chi^2$/ndof\\
\hline
5(a) & $\Upsilon(1S)$ & $24.00\pm1.00$ & $12.50\pm1.00$ & $56.89/11$ & $1.60\pm0.10$ & $9.50\pm2.00$  & $21.67/11$\\
     & $\Upsilon(2S)$ & $26.00\pm2.00$ & $12.70\pm1.00$ & $9.75/6$   & $1.75\pm0.10$ & $9.70\pm2.00$  & $7.50/6$\\
     & $\Upsilon(3S)$ & $28.00\pm2.00$ & $12.90\pm1.00$ & $14.97/6$  & $1.85\pm0.10$ & $10.50\pm2.00$ & $7.43/6$\\
5(b) & $\Upsilon(1S)$ & $27.00\pm2.00$ & $12.70\pm1.00$ & $39.17/3$  & $1.65\pm0.10$ & $11.00\pm2.00$ & $9.54/3$\\
     & $\Upsilon(2S)$ & $28.00\pm2.00$ & $12.80\pm1.00$ & $4.76/3$   & $1.90\pm0.10$ & $11.50\pm2.00$ & $4.53/3$\\
     & $\Upsilon(3S)$ & $29.00\pm2.00$ & $12.90\pm1.00$ & $2.73/3$   & $2.00\pm0.10$ & $12.00\pm2.00$ & $2.90/3$\\
5(c) & $\Upsilon(1S)$ & $29.00\pm2.00$ & $12.90\pm1.00$ & $6.60/3$   & $2.20\pm0.20$ & $16.00\pm2.00$ & $5.39/3$\\
     & $\Upsilon(2S)$ & $33.00\pm2.00$ & $13.00\pm1.00$ & $6.64/-$   & $2.50\pm0.20$ & $16.30\pm2.00$ & $6.69/-$\\
     & $\Upsilon(3S)$ & $34.00\pm2.00$ & $13.20\pm1.00$ & $11.25/-$  & $2.60\pm0.20$ & $16.50\pm2.00$ & $11.50/-$\\
5(d) & $2.0<y<2.5$ & $31.00\pm2.00$ & $15.00\pm1.00$ & $11.65/21$ & $2.30\pm0.20$ & $19.00\pm2.00$ & $7.17/21$\\
     & $2.5<y<3.0$ & $30.50\pm2.00$ & $15.50\pm1.00$ & $9.45/21$  & $2.20\pm0.20$ & $20.00\pm2.00$ & $9.48/21$\\
     & $3.0<y<3.5$ & $30.00\pm2.00$ & $16.00\pm1.00$ & $7.89/21$  & $2.10\pm0.20$ & $21.00\pm2.00$ & $6.39/21$\\
     & $3.5<y<4.0$ & $29.50\pm2.00$ & $16.50\pm1.00$ & $10.06/18$ & $1.95\pm0.20$ & $23.00\pm2.00$ & $8.07/18$\\
     & $4.0<y<4.5$ & $28.50\pm2.00$ & $17.50\pm1.00$ & $9.64/12$  & $1.80\pm0.20$ & $25.00\pm2.00$ & $7.98/12$\\
5(e) & $2.0<y<2.5$ & $36.00\pm2.00$ & $15.20\pm1.00$ & $15.98/21$ & $2.60\pm0.20$ & $20.00\pm2.00$ & $13.23/21$\\
     & $2.5<y<3.0$ & $35.50\pm2.00$ & $15.70\pm1.00$ & $8.34/21$  & $2.50\pm0.20$ & $21.00\pm2.00$ & $10.26/21$\\
     & $3.0<y<3.5$ & $35.20\pm2.00$ & $16.20\pm1.00$ & $12.84/21$ & $2.40\pm0.20$ & $22.00\pm2.00$ & $9.34/21$\\
     & $3.5<y<4.0$ & $34.30\pm1.00$ & $17.00\pm0.50$ & $10.64/18$ & $2.10\pm0.20$ & $24.00\pm2.00$ & $8.31/18$\\
     & $4.0<y<4.5$ & $31.00\pm1.00$ & $17.50\pm0.50$ & $14.06/12$ & $1.90\pm0.20$ & $26.00\pm2.00$ & $11.34/12$\\
5(f) & $2.0<y<2.5$ & $40.00\pm2.00$ & $15.50\pm1.00$ & $14.73/21$ & $2.70\pm0.20$ & $21.00\pm2.00$ & $15.45/21$\\
     & $2.5<y<3.0$ & $39.50\pm2.00$ & $16.00\pm1.00$ & $10.85/21$ & $2.60\pm0.20$ & $22.00\pm2.00$ & $7.95/21$\\
     & $3.0<y<3.5$ & $39.00\pm2.00$ & $16.50\pm1.00$ & $12.37/21$ & $2.50\pm0.20$ & $23.00\pm2.00$ & $6.78/21$\\
     & $3.5<y<4.0$ & $38.50\pm2.00$ & $17.80\pm1.00$ & $14.56/18$ & $2.30\pm0.20$ & $25.00\pm2.00$ & $8.31/18$\\
     & $4.0<y<4.5$ & $37.00\pm2.00$ & $18.30\pm1.00$ & $11.37/12$ & $2.10\pm0.20$ & $26.50\pm2.00$ & $7.53/12$\\
\hline
6(a) & $2.0<y<2.5$ & $32.00\pm2.00$ & $15.50\pm1.00$ & $14.23/21$ & $2.35\pm0.20$ & $22.00\pm2.00$ & $9.12/21$\\
     & $2.5<y<3.0$ & $31.50\pm2.00$ & $16.00\pm1.00$ & $7.53/21$  & $2.25\pm0.20$ & $23.00\pm2.00$ & $14.76/21$\\
     & $3.0<y<3.5$ & $31.00\pm2.00$ & $16.50\pm1.00$ & $12.33/21$ & $2.15\pm0.10$ & $24.00\pm2.00$ & $7.44/21$\\
     & $3.5<y<4.0$ & $30.50\pm2.00$ & $17.00\pm1.00$ & $8.72/18$  & $2.05\pm0.10$ & $25.00\pm2.00$ & $9.75/18$\\
     & $4.0<y<4.5$ & $29.00\pm2.00$ & $17.50\pm0.50$ & $13.73/12$ & $1.85\pm0.10$ & $27.00\pm2.00$ & $12.03/12$\\
6(b) & $2.0<y<2.5$ & $38.00\pm2.00$ & $16.00\pm1.00$ & $16.54/21$ & $2.70\pm0.20$ & $24.00\pm2.00$ & $10.98/21$\\
     & $2.5<y<3.0$ & $37.50\pm2.00$ & $16.50\pm1.00$ & $7.14/21$  & $2.60\pm0.10$ & $25.00\pm2.00$ & $12.54/21$\\
     & $3.0<y<3.5$ & $37.00\pm2.00$ & $17.00\pm1.00$ & $11.25/21$ & $2.50\pm0.10$ & $26.00\pm2.00$ & $8.34/21$\\
     & $3.5<y<4.0$ & $36.50\pm2.00$ & $17.20\pm0.50$ & $10.31/18$ & $2.30\pm0.10$ & $27.50\pm2.00$ & $9.51/18$\\
     & $4.0<y<4.5$ & $35.00\pm2.00$ & $17.60\pm0.50$ & $14.24/12$ & $2.00\pm0.10$ & $28.50\pm2.00$ & $5.67/12$\\
6(c) & $2.0<y<2.5$ & $44.00\pm3.00$ & $16.50\pm1.00$ & $15.89/21$ & $2.90\pm0.10$ & $25.00\pm2.00$ & $9.48/21$\\
     & $2.5<y<3.0$ & $43.00\pm3.00$ & $17.00\pm1.00$ & $7.53/21$  & $2.80\pm0.20$ & $25.50\pm2.00$ & $10.68/21$\\
     & $3.0<y<3.5$ & $42.00\pm3.00$ & $17.50\pm1.00$ & $13.59/21$ & $2.70\pm0.10$ & $26.50\pm2.00$ & $7.53/21$\\
     & $3.5<y<4.0$ & $41.00\pm3.00$ & $18.00\pm1.00$ & $12.42/18$ & $2.50\pm0.10$ & $28.00\pm2.00$ & $7.34/18$\\
     & $4.0<y<4.5$ & $38.00\pm3.00$ & $18.50\pm1.00$ & $13.59/12$ & $2.23\pm0.10$ & $29.00\pm2.00$ & $8.31/12$\\
6(d) & $1.5<y<2.0$ & $34.00\pm2.00$ & $15.20\pm1.00$ & $5.35/4$   & $2.45\pm0.10$ & $22.00\pm2.00$ & $3.85/4$\\
     & $2.0<y<2.5$ & $33.00\pm2.00$ & $15.70\pm1.00$ & $4.79/4$   & $2.35\pm0.10$ & $23.00\pm2.00$ & $2.20/4$\\
     & $2.5<y<3.0$ & $32.50\pm2.00$ & $16.00\pm1.00$ & $4.25/4$   & $2.25\pm0.10$ & $24.00\pm2.00$ & $2.69/4$\\
     & $3.0<y<3.5$ & $32.00\pm2.00$ & $16.50\pm1.00$ & $6.72/4$   & $2.15\pm0.10$ & $25.00\pm2.00$ & $3.70/4$\\
     & $3.5<y<4.0$ & $31.50\pm2.00$ & $17.00\pm1.00$ & $7.40/3$   & $2.05\pm0.10$ & $26.00\pm2.00$ & $4.40/3$\\
6(e) & $2.0<y<2.5$ & $37.00\pm3.00$ & $16.00\pm1.00$ & $16.58/21$ & $2.60\pm0.10$ & $26.00\pm2.00$ & $19.66/21$\\
     & $2.5<y<3.0$ & $36.00\pm3.00$ & $16.50\pm1.00$ & $10.31/22$ & $2.50\pm0.10$ & $27.00\pm2.00$ & $14.83/22$\\
     & $3.0<y<3.5$ & $35.00\pm2.00$ & $17.00\pm0.50$ & $10.28/21$ & $2.40\pm0.10$ & $28.00\pm2.00$ & $13.91/21$\\
     & $3.5<y<4.0$ & $34.50\pm2.00$ & $17.50\pm0.50$ & $13.42/19$ & $2.30\pm0.10$ & $29.00\pm2.00$ & $12.85/19$\\
     & $4.0<y<4.5$ & $34.30\pm1.00$ & $17.70\pm0.50$ & $9.64/13$  & $2.25\pm0.10$ & $29.50\pm2.00$ & $8.53/13$\\
6(f) & $2.0<y<2.5$ & $41.00\pm3.00$ & $16.50\pm1.00$ & $20.71/21$ & $2.80\pm0.10$ & $29.00\pm2.00$ & $12.23/21$\\
     & $2.5<y<3.0$ & $40.00\pm2.00$ & $17.00\pm0.50$ & $18.20/22$ & $2.70\pm0.10$ & $30.00\pm2.00$ & $24.00/22$\\
     & $3.0<y<3.5$ & $39.50\pm2.00$ & $17.20\pm0.50$ & $7.92/21$  & $2.65\pm0.10$ & $30.50\pm2.00$ & $8.31/21$\\
     & $3.5<y<4.0$ & $39.20\pm1.00$ & $17.40\pm0.50$ & $16.12/19$ & $2.60\pm0.10$ & $31.00\pm2.00$ & $14.11/19$\\
     & $4.0<y<4.5$ & $39.00\pm1.00$ & $17.80\pm0.50$ & $14.17/13$ & $2.50\pm0.10$ & $32.00\pm2.00$ & $10.81/13$\\
6(g) & $2.0<y<2.5$ & $47.00\pm2.00$ & $17.00\pm1.00$ & $16.98/21$ & $3.20\pm0.10$ & $30.00\pm2.00$ & $12.63/21$\\
     & $2.5<y<3.0$ & $46.00\pm2.00$ & $17.50\pm1.00$ & $19.68/22$ & $3.10\pm0.10$ & $31.00\pm2.00$ & $10.16/22$\\
     & $3.0<y<3.5$ & $45.50\pm2.00$ & $17.70\pm0.50$ & $14.69/21$ & $3.00\pm0.10$ & $32.00\pm2.00$ & $10.40/21$\\
     & $3.5<y<4.0$ & $45.30\pm2.00$ & $18.50\pm1.00$ & $15.60/19$ & $2.90\pm0.10$ & $33.00\pm2.00$ & $11.71/19$\\
     & $4.0<y<4.5$ & $45.00\pm2.00$ & $18.80\pm0.50$ & $14.66/13$ & $2.80\pm0.10$ & $34.00\pm2.00$ & $8.86/13$\\
\hline
\end{tabular}}
\end{center}


\newpage

{\bf Appendix 2: Figures of parameters and some discussions.}
\\

\begin{figure*}[!htb]
\begin{center}
\includegraphics[width=9.5cm]{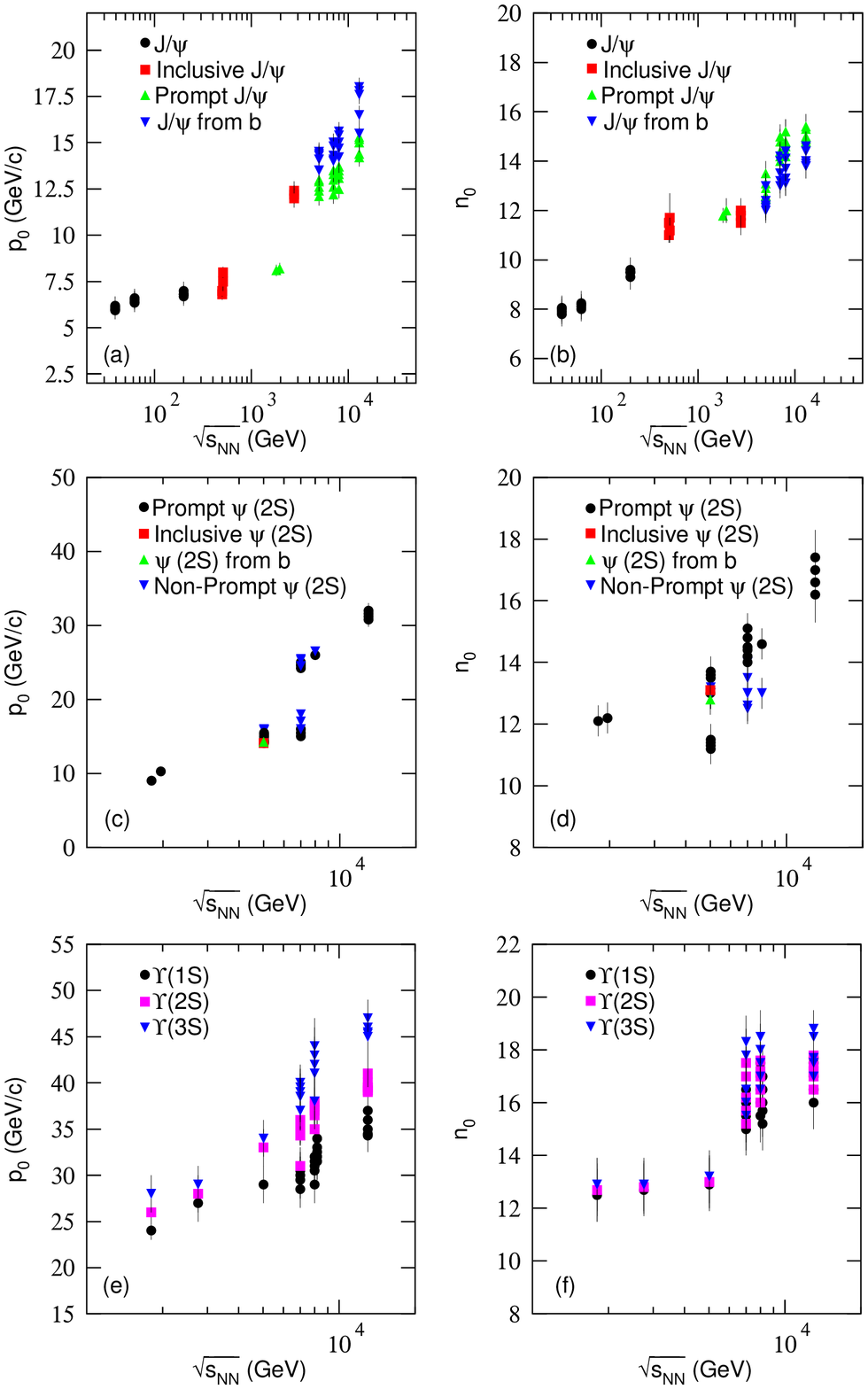}
\end{center}
{\small Fig. A1. Same as Fig. 7, but showing the dependences of
(a)(c)(e) $p_0$ and (b)(d)(f) $n_0$ on $\sqrt{s_{NN}}$ (or
$\sqrt{s}$) for (a)(b) $J/\psi$, (c)(d) $\psi(2S)$, and (e)(f)
$\Upsilon(nS, n=1,2,3)$.}
\end{figure*}

\begin{figure*}[!htb]
\begin{center}
\includegraphics[width=9.5cm]{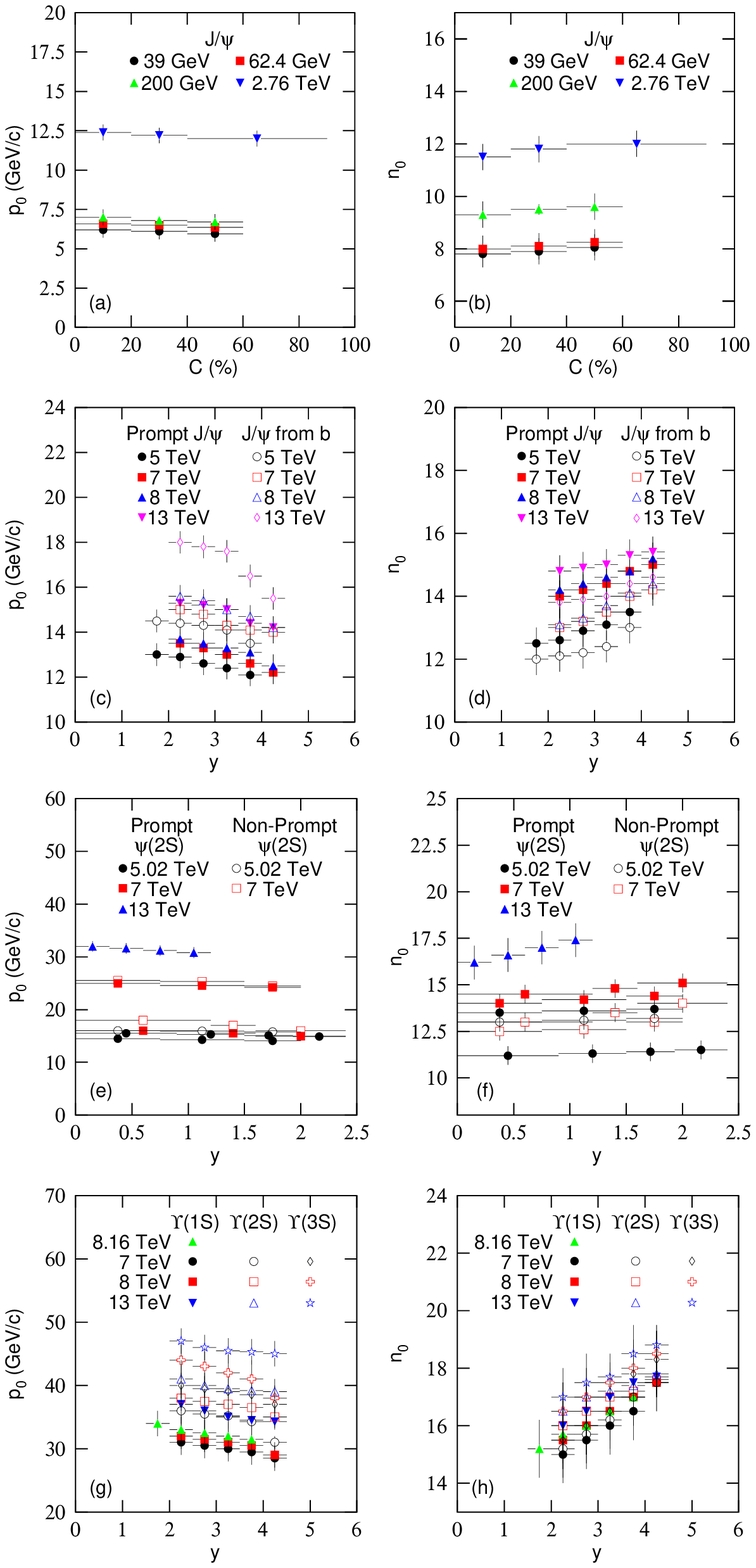}
\end{center}
{\small Fig. A2. Same as Fig. 7, but showing the dependences of
(a)(c)(e)(g) $p_0$ and (b)(d)(f)(h) $n_0$ on (a)(b) $C$ and
(c)--(h) $y$ for (a)--(d) $J/\psi$, (e)(f) $\psi(2S)$, and (g)(h)
$\Upsilon(nS, n=1,2,3)$.}
\end{figure*}

\begin{figure*}[!htb]
\begin{center}
\includegraphics[width=9.5cm]{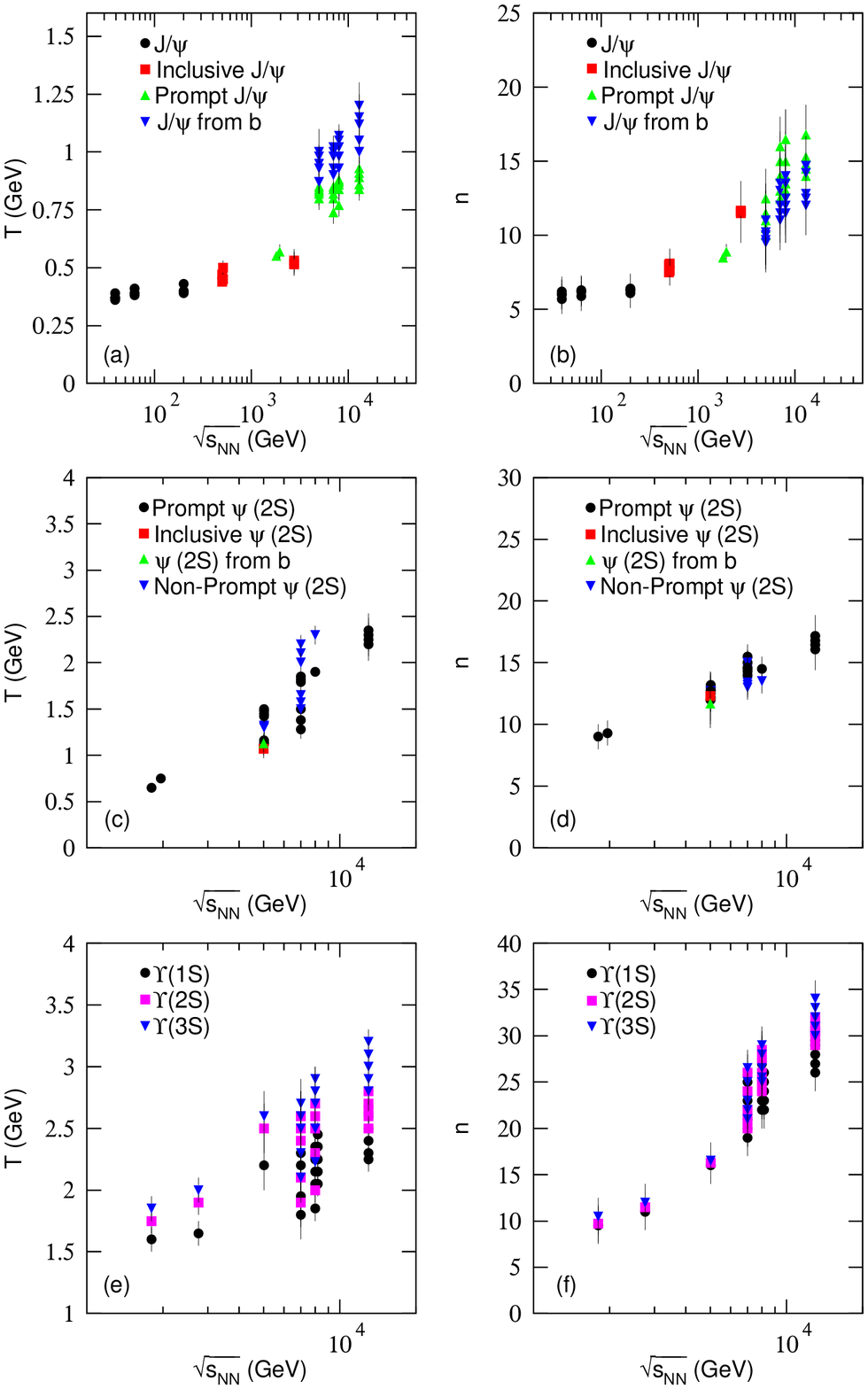}
\end{center}
{\small Fig. A3. Same as Fig. 7, but showing the dependences of
(a)(c)(e) $T$ and (b)(d)(f) $n$ on $\sqrt{s_{NN}}$ (or $\sqrt{s}$)
for (a)(b) $J/\psi$, (c)(d) $\psi(2S)$, and (e)(f) $\Upsilon(nS,
n=1,2,3)$, where $T$ is not the initial temperature, but an
effective temperature parameter in the Tsallis-Levy function.}
\end{figure*}

\begin{figure*}[!htb]
\begin{center}
\includegraphics[width=9.5cm]{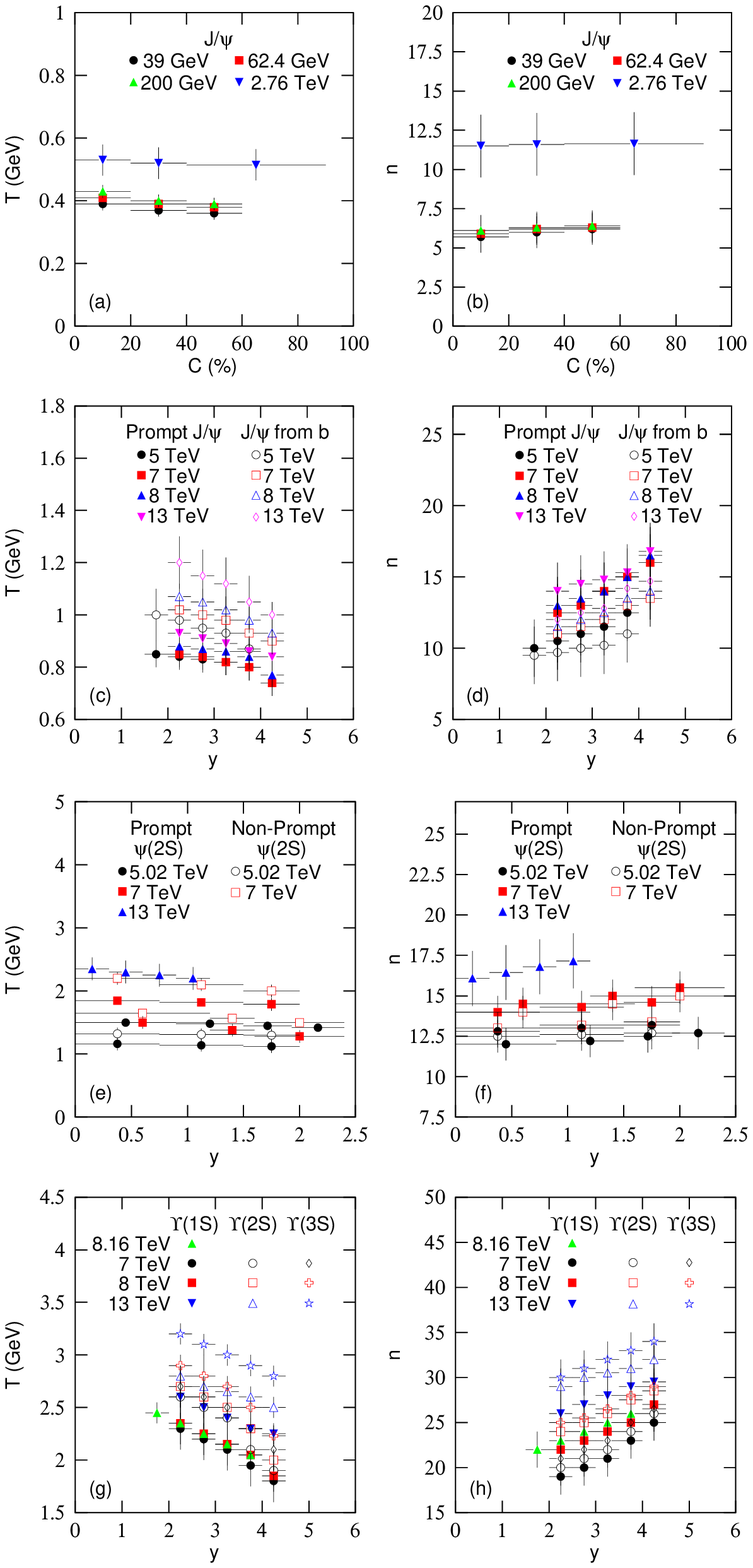}
\end{center}
{\small Fig. A4. Same as Fig. 7, but showing the dependences of
(a)(c)(e)(g) $T$ and (b)(d)(f)(h) $n$ on (a)(b) $C$ and (c)--(h)
$y$ for (a)--(d) $J/\psi$, (e)(f) $\psi(2S)$, and (g)(h)
$\Upsilon(nS, n=1,2,3)$, where $T$ is not the initial temperature,
but an effective temperature parameter in the Tsallis-Levy
function.}
\end{figure*}

\begin{multicols}{2}

Figure A1 is the same as Fig. 7, but showing the dependences of
(a)(c)(e) $p_0$ and (b)(d)(f) $n_0$ on $\sqrt{s_{NN}}$ (or
$\sqrt{s}$) for (a)(b) $J/\psi$, (c)(d) $\psi(2S)$, and (e)(f)
$\Upsilon(nS, n=1,2,3)$. The different symbols represent the
parameter values derived from free parameters extracted from Figs.
1--6 and listed in Tables A1 and A2. One can see that $p_0$ and
$n_0$ increases with the increase of collision energy and particle
mass.

Figure A2 is the same as Fig. 7, but showing the dependences of
(a)(c)(e)(g) $p_0$ and (b)(d)(f)(h) $n_0$ on (a)(b) centrality $C$
and (c)--(h) rapidity $y$ for (a)--(d) $J/\psi$, (e)(f)
$\psi(2S)$, and (g)(h) $\Upsilon(nS, n=1,2,3)$. One can see that
$p_0$ ($n_0$) increases (decreases) slightly with the increase of
event centrality from peripheral to central collisions, and
decreases (increase) (slightly) with the increase of rapidity from
mid-rapidity to forward rapidity. Meanwhile, $p_0$ ($n_0$)
increases with the increases of collision energy and particle
mass.

Figure A3 is the same as Fig. 7, but showing the dependences of
(a)(c)(e) $T$ and (b)(d)(f) $n$ on $\sqrt{s_{NN}}$ (or $\sqrt{s}$)
for (a)(b) $J/\psi$, (c)(d) $\psi(2S)$, and (e)(f) $\Upsilon(nS,
n=1,2,3)$. One can see that $T$ and $n$ increases with the
increase of collision energy and particle mass.

Figure A4 is the same as Fig. 7, but showing the dependences of
(a)(c)(e)(g) $T$ and (b)(d)(f)(h) $n$ on (a)(b) centrality $C$ and
(c)--(h) rapidity $y$ for (a)--(d) $J/\psi$, (e)(f) $\psi(2S)$,
and (g)(h) $\Upsilon(nS, n=1,2,3)$. One can see that $T$ ($n$)
increases (decreases) slightly with the increase of event
centrality from peripheral to central collisions, and decreases
(increase) (slightly) with the increase of rapidity from
mid-rapidity to forward rapidity. Meanwhile, $T$ ($n$) increases
with the increases of collision energy and particle mass.

The tendency of $p_0$ ($T$) and $n_0$ ($n$) with collision energy
are also explained by more violent collision at higher energy.
Both $p_0$ ($T$) and $n_0$ ($n$) increase with the increase of
collision energy. This means that $p_0$ ($T$) and $n_0$ ($n$) are
positively correlative at different energies. Meanwhile, for a
given $p_T$ spectrum or in given collisions, an increase in $p_0$
($T$) is concomitant with a decrease in $n_0$ ($n$). This means
that $p_0$ ($T$) and $n_0$ ($n$) are negatively correlative in
given collisions (at given energy). There are correlations between
$p_0$ ($T$) and $n_0$ ($n$) when we determine these parameters.

The correlation between $p_0$ ($T$) and $n_0$ ($n$) is similar to
that between kinetic freeze-out temperature and transverse flow
velocity~\cite{56,57} which also show positive correlation at
different energies and negative correlation in given spectrum. If
$p_0$ ($T$) is similar to kinetic freeze-out temperature, $n_0$
($n$) should be similar to transverse flow velocity. Meanwhile,
the results obtained in this work are in agreement with our recent
work~\cite{58}, which shows mass-dependent parameters. In
particular, with the increase of particle mass, $\langle
p_T\rangle$, $T_i$, $p_0$, and $n_0$ increase.

\end{multicols}

{\small
\begin{thebibliography}{99}
\setlength{\itemsep}{-1pt}

\bibitem{1}
J. K. Nayak, J. Alam, S. Sarkar, B. Sinha, ``Measuring initial
temperature through a photon to dilepton ratio in heavy-ion
collisions," {\it Journal of Physics G}, vol. 35, article 104161,
2008.

\bibitem{1a}
A. Adare et al. [PHENIX Collaboration], ``Enhanced production of
direct photons in Au+Au collisions at $\sqrt{s_{NN}}=200$ GeV and
implications for the initial temperature," {\it Physical Review
Letters}, vol. 104, article 132301, 2010.

\bibitem{1aa}
M. Csan{\'a}d and I. M{\'a}jer, ``Initial temperature and EoS of
quark matter via direct photons," {\it Physics of Particles and
Nuclei Letters}, vol 8, pp. 1013--1015, 2011.

\bibitem{1aaa}
M. Csan{\'a}d and I. M{\'a}jer, ``Equation of state and initial
temperature of quark gluon plasma at RHIC," {\it Central European
Journal of Physics}, vol. 10, pp. 850--857, 2012.

\bibitem{1aaaa}
R. A. Soltz, I. Garishvili, M. Cheng, B. Abelev, A. Glenn, J.
Newby, L. A. L. Levy, and S. Pratt, ``Constraining the initial
temperature and shear viscosity in a hybrid hydrodynamic model of
$\sqrt{s_{NN}}=200$ GeV Au+Au collisions using pion spectra,
elliptic flow, and femtoscopic radii," {\it Physical Review C},
vol. 87, article 044901, 2013.

\bibitem{1b}
F.-H. Liu and J.-S. Li, ``Isotopic production cross section of
fragments in $^{56}$Fe+p and $^{136}$Xe($^{124}$Xe)+Pb reactions
over an energy range from $300A$ to $1500A$ MeV," {\it Physical
Review C}, vol. 78, article 044602, 2008.

\bibitem{2}
F.-H. Liu, ``Unified description of multiplicity distributions of
final-state particles produced in collisions at high energies,"
{\it Nuclear Physics A}, vol. 810, pp. 159--172, 2008.

\bibitem{3}
F.-H. Liu, Y.-Q. Gao, T. Tian, B.-C. Li, ``Unified description of
transverse momentum spectrums contributed by soft and hard
processes in high-energy nuclear collisions," {\it The European
Physical Journal A}, vol. 50, article 94, 2014.

\bibitem{4a}
R. Hagedorn, ``Multiplicities, $p_T$ distributions and the
expected hadron $\longrightarrow$ quark-gluon phase transition,"
{\it La Rivista del Nuovo Cimento}, vol. 6, no. 10, pp. 1--50
(1983).

\bibitem{4b}
B. Abelev et al. [ALICE Collaboration], ``Production of
$\Sigma(1385)^\pm$ and $\Sigma(1530)^0$ in proton-proton
collisions at $\sqrt{s}=7$ TeV," {\it The European Physical
Journal C}, vol. 75, article 1, 2015.

\bibitem{5a}
C. Tsallis, ``Possible generalization of Boltzmann-Gibbs
statistics," {\it Journal of Statistical Physics}, vol. 52, pp.
479--487, 1988.

\bibitem{5b}
B. I. Abelev et al. [STAR Collaboration], ``Strange particle
production in p+p collisions at $\sqrt{s}=200$ GeV," {\it Physical
Review C}, vol. 75, article 064901, 2007.

\bibitem{27}
L. Adamczyk et al. [STAR Collaboration], ``Energy dependence of
$J/\psi$ production in Au+Au collisions at $\sqrt{s_{NN}}=39$,
62.4 and 200 GeV," {\it Physics Letters B}, vol. 711, pp. 13--20,
2017.

\bibitem{28}
J. Adam et al. [STAR Collaboration], ``Measurements of the
transverse-momentum-dependent cross sections of $J/\psi$
production at mid-rapidity in proton+proton collisions at
$\sqrt{s}=510$ and 500 GeV with the STAR detector," {\it Physical
Review D}, vol. 100, article 052009, 2019.

\bibitem{29a}
F. Abe et al. [CDF Collaboration], ``$J/\psi$ and $\psi(2S)$
production in $p\overline{p}$ collisions at $\sqrt{s}=1.8$ TeV,"
{\it Physical Review Letters}, vol. 79, pp. 572--577, 1997.

\bibitem{29b}
D. Acosta et al. [CDF Collaboration], ``Measurement of the
$J/\psi$ and $b$--hadron production cross sections in
$p\overline{p}$ collisions at $\sqrt{s}=1960$ GeV," {\it Physical
Review D}, vol. 71, article 032001, 2005.

\bibitem{41}
D. Acosta et al. [CDF Collaboration], ``$\Upsilon$ production and
polarization in $p\overline{p}$ collisions at $\sqrt{s}=1.8$ TeV,"
{\it Physical Review Letters}, vol. 88, article 161802, 2002.

\bibitem{30}
J. Adam et al. [ALICE Collaboration], ``Differential studies of
inclusive $J/\psi$ and $\psi(2S)$ production at forward rapidity
in Pb-Pb collisions at $\sqrt{s_{NN}}=2.76$ TeV," {\it Journal of
High Eenergy Physics}, vol. 16, no. 05, article 179, 2016.

\bibitem{31}
R. Aaij et al. [LHCb Collaboration], ``Study of $J/\psi$
production and cold nuclear matter effects in $p$Pb collisions at
$\sqrt{s_{NN}}=5$ TeV," {\it Journal of High Energy Physics}, vol.
14, no. 02, article 072, 2014.

\bibitem{32}
R. Aaij et al. [LHCb Collaboration], ``Measurement of $J/\psi$
production in $pp$ collisions at $\sqrt{s}=7$ TeV," {\it The
European Physical Journal C}, vol. 71, article 1645, 2011.

\bibitem{33}
R. Aaij et al. [LHCb Collaboration], ``Production of $J/\psi$ and
$\Upsilon$ mesons in $pp$ collisions at $\sqrt{s}=8$ TeV," {\it
Journal of High Energy Physics}, vol. 13, no. 06, article 064,
2013.

\bibitem{34}
R. Aaij et al. [LHCb Collaboration], ``Measurement of forward
$J/\psi$ production cross-sections in $pp$ collisions at
$\sqrt{s}=13$ TeV," {\it Journal of High Energy Physics}, vol. 11,
no. 05, article 172, 2015.

\bibitem{35}
R. Aaij et al. [LHCb Collaboration], ``Study of $\psi(2S)$
production and cold nuclear matter effects in $p$Pb collisions at
$\sqrt{s_{NN}}=5$ TeV," {\it Journal of High Energy Physics}, vol.
16, no. 03, article 133, 2016.

\bibitem{42}
R. Aaij et al. [LHCb Collaboration], ``Measurement of $\Upsilon$
production in $pp$ collisions at $\sqrt{s}=2.76$ TeV," {\it The
European Physical Journal C}, vol. 74, article 2835, 2014.

\bibitem{44}
R. Aaij et al. [LHCb Collaboration], ``Forward production of
$\Upsilon$ mesons in $pp$ collisions at $\sqrt{s}=7$ and 8 TeV,"
{\it Journal of High Energy Physics}, vol. 15, no. 11, article
103, 2015.

\bibitem{45}
R. Aaij et al. [LHCb Collaboration], ``Study of $\Upsilon$
production in $p$Pb collisions at $\sqrt{s_{NN}}=8.16$ TeV," {\it
Journal of High Energy Physics}, vol. 18, no. 11, article 194,
2018.

\bibitem{46}
R. Aaij et al. [LHCb Collaboration], ``Measurement of $\Upsilon$
production in $pp$ collisions at $\sqrt{s}=13$ TeV," {\it Journal
of High Energy Physics}, vol. 18, no. 07, article 134, 2018.

\bibitem{36}
M. Aaboud et al. [ATLAS Collaboration], ``Measurement of
quarkonium production in proton-lead and proton-proton collisions
at 5.02 TeV with the ATLAS detector," {\it The European Physical
Journal C}, vol. 78, article 171, 2018.

\bibitem{38b}
G. Aad et al. [ATLAS Collaboration], ``Measurement of the
production cross-section of $\psi(2S)\rightarrow J/\psi
(\rightarrow \mu^+\mu^-) \pi^+\pi^-$ in $pp$ collisions at
$\sqrt{s}=7$ TeV at ATLAS," {\it Journal of High Energy Physics},
vol. 14, no. 09, article 079, 2014.

\bibitem{39}
M. Aaboud et al. [ATLAS Collaboration], ``Measurement of
$\psi(2S)$ and $X(3872)\rightarrow J/\psi\pi^+\pi^-$ production in
$pp$ collisions at $\sqrt{s}=8$ TeV with the ATLAS detector," {\it
Journal of High Energy Physics}, vol. 17, no. 01, article 117,
2017.

\bibitem{37}
A. M. Sirunyan et al. [CMS Collaboration], ``Measurement of prompt
$\psi(2S)$ production cross sections in proton-lead and
proton-proton collisions at $\sqrt{s_{NN}}=5.02$ TeV," {\it
Physics Letters B}, vol. 790, pp. 509--532, 2019.

\bibitem{38a}
S. Chatrchyan et al. [CMS collaboration], ``$J/\psi$ and
$\psi(2S)$ production in $pp$ collisions at $\sqrt{s_{NN}}=7$
TeV," {\it Journal of High Energy Physics}, vol. 12, no. 02,
article 011, 2012.

\bibitem{40}
A. M. Sirunyan et al. [CMS Collaboration], ``Measurement of
quarkonium production cross sections in $pp$ collisions at
$\sqrt{s}=13$ TeV," {\it Physics Letters B}, vol. 780, pp.
251--272, 2018.

\bibitem{43}
A. M. Sirunyan et al. [CMS Collaboration], ``Measurement of
nuclear modification factors of $\Upsilon(1S)$, $\Upsilon(2S)$,
and $\Upsilon(3S)$ mesons in PbPb collisions at
$\sqrt{s_{NN}}=5.02$ TeV," {\it Physics Letters B}, vol. 790, pp.
270--293, 2019.

\bibitem{6}
E. K. G. Sarkisyan and A. S. Sakharov, ``On similarities of bulk
observables in nuclear and particle collisions,"
CERN-PH-TH-2004-213, arXiv:hep-ph/0410324, 2004.

\bibitem{7}
E. K. G. Sarkisyan and A. S. Sakharov, ``Multihadron production
features in different reactions," {\it AIP Conference
Proceedings}, vol. 828, pp. 35--41, 2006.

\bibitem{8}
E. K. G. Sarkisyan and A. S. Sakharov, ``Relating multihadron
production in hadronic and nuclear collisions," {\it The European
Physical Journal C}, vol. 70, pp. 533--541, 2010.

\bibitem{9}
E. K. G. Sarkisyan, A. N. Mishra, R. Sahoo, and A. S. Sakharov,
``Multihadron production dynamics exploring the energy balance in
hadronic and nuclear collisions," {\it Physical Review D}, vol.
93, article 054046, 2016.

\bibitem{10}
E. K. G. Sarkisyan, A. N. Mishra, R. Sahoo, and A. S. Sakharov,
``Centrality dependence of midrapidity density from GeV to TeV
heavy-ion collisions in the effective-energy universality picture
of hadroproduction," {\it Physical Review D}, vol. 94, article
011501(R), 2016.

\bibitem{11}
W. Kittel and E. A. De Wolf, Soft Multihadron Dynamics. p. 652.
World Scientific, Singapore, 2005.

\bibitem{12}
R. Nouicer, ``Similarity of initial states in A+A and p+p
collisions in constituent quarks framework," {\it AIP Conference
Proceedings}, vol. 828, pp. 11--16, 2006.

\bibitem{13}
R. Nouicer for the PHOBOS Collaboration, ``Systematic of global
observables in Cu+Cu and Au+Au collisions at RHIC energies," {\it
AIP Conference Proceedings}, vol. 842, pp. 86--88, 2006.

\bibitem{14}
R. Nouicer, ``Charged particles multiplicities in A+A and p+p
collisions in the constituent quarks framework," {\it The European
Physical Journal C}, vol. 49, pp. 281--286, 2007.

\bibitem{15}
J. F. Grosse-Oetringhaus and K. Reygers, ``Charged particle
multiplicity in proton-proton collisions," {\it Journal of Physics
G}, vol. 37, article 083001, 2020.

\bibitem{16}
A. N. Mishra, A. Ortiz, and G. Pai{\'c}, ``Intriguing similarities
of high-$p_T$ particle production between pp and A-A Collisions,"
{\it Physical Review C}, vol. 99, article 034911, 2019.

\bibitem{17}
K. Aamodt et al. [ALICE Collaboration], ``Transverse momentum
spectra of charged particles in proton-proton collisions at
$\sqrt{s}=900$ GeV with ALICE at the LHC," {\it Physics Letters
B}, vol. 693, pp. 53--68, 2020.

\bibitem{18}
A. De Falco for the ALICE Collaboration, ``Vector meson production
in pp collisions at $\sqrt{s}=7$ TeV, measured with the ALICE
detector," {\it Journal of Physics G}, vol. 38, article 124083,
2011.

\bibitem{19}
A. Adare et al. [PHENIX Collaboration], ``Nuclear modification
factors of $\phi$ mesons in d+Au, Cu+Cu, and Au+Au collisions at
$\sqrt{s_{NN}}=200$ GeV," {\it Physical Review C}, vol. 83,
article 024909, 2011.

\bibitem{20}
B. Abelev et al. [ALICE Collaboration], ``Light vector meson
production in pp collisions at $\sqrt{s}=7$ TeV," {\it Physics
Letters B}, vol. 710, pp. 557--568, 2012.

\bibitem{21}
B. Abelev et al. [ALICE Collaboration], ``Inclusive $J/\psi$
production in pp collisions at $\sqrt{s}=2.76$ TeV," {\it Physics
Letters B}, vol. 718, pp. 295--306, 2012.

\bibitem{22}
B. Abelev et al. [ALICE Collaboration], ``Heavy flavour decay muon
production at forward rapidity in proton-proton collisions at
$\sqrt{s}=7$ TeV," {\it Physics Letters B}, vol. 708, pp.
265--275, 2012.

\bibitem{23}
I. Lakomov for the ALICE collaboration, ``Event activity
dependence of inclusive $J/\psi$ production in p-Pb collisions at
$\sqrt{s_{NN}}=5.02$ TeV with ALICE at the LHC," {\it Nuclear
Physics A}, vol. 931, pp. 1179--1183, 2014.

\bibitem{24}
L. G. Gutay, A. S. Hirsch, C. Pajares, R. P. Scharenberg, and B.
K. Srivastava, ``De-confinement in small systems: Clustering of
color sources in high multiplicity $\bar pp$ collisions at
$\sqrt{s_{NN}}=1.8$ TeV," {\it International Journal of Modern
Physics E}, vol. 24, article 1550101, 2015.

\bibitem{25}
A. S. Hirsch, C. Pajares, R. P. Scharenberg, and B. K. Srivastava,
``De-confinement in high multiplicity proton-proton collisions at
LHC energies," {\it Physical Review D}, vol. 100, article 114040,
2019.

\bibitem{26}
P. Sahoo, S. De, S. K. Tiwari, and R. Sahoo, ``Energy and
centrality dependent study of deconfinement phase transition in a
color string percolation approach at RHIC energies," {\it The
European Physical Journal A}, vol. 54, article 136, 2018.

\bibitem{55a}
N. Sarkar, P. Deb, and P. Ghosh, ``Finite size effect on
thermodynamics of hadron gas in high-multiplicity events of
proton-proton collisions at the LHC," arXiv:1905.06532 [hep-ph],
2019.

\bibitem{55b}
K. Shen, G. G. Barnaf{\"o}ldi, and T. S. Bir{\'o}, Hadron spectra
parameters within the non-extensive approach," {\it Universe}, vol
5, article 122, 2019.

\bibitem{55c}
R. Rath, A. Khuntia, and R. Sahoo, ``System size and multiplicity
dependence of chemical freeze-out parameters at the Large Hadron
Collider energies," arXiv:1905.07959 [hep-ph], 2019.

\bibitem{55d}
R. Sahoo, ``Possible formation of QGP-droplets in proton-proton
collisions at the CERN Large Hadron Collider," {\it AAPPS
Bulletin}, vol. 29, no. 4, pp. 16--21, 2019.

\bibitem{56}
M. Waqas and B.-C. Li, ``Kinetic freeze-out temperature and
transverse flow velocity in Au-Au collisions at RHIC-BES
energies," {\it Advances in High Energy Physics}, vol. 2020,
article 1787183, 2020.

\bibitem{57}
P.-P. Yang, M.-Y.  Duan, F.-H. Liu, and R. Sahoo, ``Multiparticle
production and initial quasi-temperature from proton induced
carbon collisions at $p_{Lab}=$ 31 GeV/$c$," {\it Advances in High
Energy Physics}, vol. 2020, article 9542196, 2020.

\bibitem{58}
Q. Wang and F.-H. Liu, ``Initial and final state temperatures of
antiproton emission sources in high energy collisions," {\it
International Journal of Theoretical Physics}, vol. 58, pp.
4119--4138, 2019.

\end{thebibliography}}
\end{document}